\documentclass{JHEP3}

\pdfoutput=1

\usepackage{graphicx}
\usepackage{amssymb}
\usepackage{slashed}
\usepackage{amsmath}
\usepackage{url}
\usepackage{epsfig}

\def\beq{\begin{equation}}
\def\eeq{\end{equation}}

\newcommand{\bea}{\begin{eqnarray}}
\newcommand{\eea}{\end{eqnarray}}

\def\gev{\rm GeV}

\def\eeqn{\end{equation}}
\newcommand\iden{\leavevmode\hbox{\small1\normalsize\kern-.33em1}}

\let\jnfont=\rm
\def\NPB#1,{{\jnfont Nucl.\ Phys.\ B }{\bf #1},}
\def\PLB#1,{{\jnfont Phys.\ Lett.\ B }{\bf #1},}
\def\EPJC#1,{{\jnfont Eur.\ Phys.\ Jour.\ C }{\bf #1},}
\def\PRD#1,{{\jnfont Phys.\ Rev.\ D }{\bf #1},}
\def\PRL#1,{{\jnfont Phys.\ Rev.\ Lett.\ }{\bf #1},}
\def\MPLA#1,{{\jnfont Mod.\ Phys.\ Lett.\ A }{\bf #1},}
\def\JPG#1,{{\jnfont J.\ Phys.\ G }{\bf #1},}
\def\CTP#1,{{\jnfont Commun.\ Theor.\ Phys.\ }{\bf #1},}
\def\JHEP#1,{{\jnfont JHEP \ }{\bf #1},}

\def\NPPS#1,{{\jnfont Nucl.\ Phys.\ Proc.\ Suppl.\ }{\bf #1},}

\title{Hadronic $b^\prime$ search at the LHC with top and W taggers}

\author{Shuo Yang$^{a}$, Ji Jiang$^{b}$, Qi-Shu Yan$^{c,d}$, Xiaoran Zhao$^{c}$
\\ $^{a}$ Physics Department, Dalian University, Dalian, 116622, P.R. China
\\ $^b$ Physics Department, Zhengzhou University, Zhengzhou, 450001, P.R China
\\$^{c}$ College of Physics Sciences, University of Chinese Academy of Sciences, Beijing 100039, P.R. China
\\ $^{d}$ Center for High-Energy Physics, Peking University, Beijing, 100871, P.R. China
\\E-mail: \email{yanqishu@ucas.ac.cn}
}

\abstract{
We study the sensitivity of a down type quark $b^{\prime}$ via process $pp\rightarrow b^{\prime}\bar{b^{\prime}} \rightarrow tW^-\bar{t}W^+$ using jet substructure methods at the LHC with the collision energy $\sqrt{s}=14$ TeV.  We consider the case that the $b^\prime$ is heavy (say from 800 GeV to 1500 GeV) and concentrate on the feasibility of the full hadronic mode. Both top tagger (the HEP top tagger) and W tagger (the CMS W-tagging) are used to reconstruct all objects in the final states. In order to suppress huge SM background events and take into account various cases with different number of boosted objects, we propose a comprehensive reconstruction procedure so as to extract the most crucial observables of the signal events.  When $b^\prime$ mass is 1 TeV, it is found that with a 200 fb$^{-1}$ dataset, the LHC may be able to detect the $b^\prime$ with a significance up to $10$ or better. With a 3000 $fb^{-1}$ dataset, the LHC may be able to probe the $b^\prime$ with a mass around up to 2 TeV, only by using the hadronic mode.}


\begin{document}

\section{Introduction}
The LHC collaborations have discovered a Higgs boson \cite{Aad:2012tfa,Chatrchyan:2012ufa}, it is quite natural to ask what will be the next discovery that could be expected for future LHC runs. Extra quarks are one of the possible signals for the new physics, which are supposed to offer solutions to the fundamental issues of the standard model on electroweak symmetry breaking and mass generation of fundamental particles \cite{DeSimone:2012fs}. The heavy bottom like quarks have been predicted in Top-Coloron model \cite{Chivukula:2013kw} and top flavor seesaw \cite{Wang:2013jwa}, non-minimal supersymmetric extentions \cite{SUSYvectorlike} and extra dimension models with warped space \cite{Casagrande:2010si}, etc.

The extra quarks are good targets for future LHC runs due to their strong interaction with the particles of the SM, especially with gluons. The LHC can be called as a gluon-gluon machine due to the large gluon fluxes in highly accelerated protons. When kinematically accessible, these extra quarks can be copiously produced at the LHC. There are quite a few phenomenological studies for the feasibility of heavy quarks at the LHC, which can be found in literatures \cite{Skiba,Holdom:2007ap,Holdom:2010fr,Holdom:2011uv,AguilarSaavedra:2009es}.  Meanwhile, signature of extra quarks are one of focus for experimental searches \cite{b'ATLAS2012,b'CMS2012,Bhattacharya:2013poa,CMSB2G,CMSB2Gb'}.

In this paper, we focus on the search of vector-like $b^{\prime}$.
\footnote{The naive fourth generation chiral quarks have mostly been ruled out by the Higgs data \cite{Djouadi:2012ae}.} Vector-like fermions do not contribute to "oblique parameters" in the leading order, and thus these
parameters do not constrain their masses. However, the mixing angles between the vector-like fermions and the SM three generations fermions are required to be small because there is no GIM mechanism to suppress the FCNC related to these vector-like fermions.
Currently, the most stringent limit on $b^{\prime}$ comes from CMS search \cite{CMSB2Gb'}. Focusing on the strong pair-production mechanism, CMS has set
lower limits between 582 and 732 GeV on the vector-like $b^{\prime}$ quark mass for various decay branching ratios at 95\% confidence level \cite{CMSB2Gb'}.
If the $b^{\prime}$ exclusively decays into a top quark and a W boson, considering the same sign lepton final state, the $b^{\prime}$ with mass below 732 GeV has been excluded at the 95\% confidence level.

It is well-known that the LHC is a top quark factory and the discovery and precision measurements of $pp\to t {\bar t}$ in all the decay modes have a great significance to test the prediction of the SM. Among all the decay modes, the signal of the semi-leptonic mode is relatively easier to pick out, which enjoys a relative larger branching fraction and smaller SM backgrounds. The dileptonic mode has the smallest branching fraction but enjoys an even cleaner backgrounds \cite{Abe:1995hr,Abachi:1995iq}. The fully hadronic mode is relatively difficult due to the large QCD and W + jets  backgrounds and large uncertainties in determining the QCD activities. Nevertheless, measuring the fully hadronic mode is an important indispensable test of the SM prediction. The success of the measurement of fully hadronic mode at the Tevatron \cite{Abazov:2006yb,Aaltonen:2007as} and the LHC \cite{Chatrchyan:2013xza} demonstrates that these detectors of hadronic machines are capable to detect the final state with a large multiplicity of jets and new physics with full hadronic final states \cite{fullhadron-ttbar-tev,fullhadron-ttbar-lhc-jetsub}. Recently, phenomenological studies using the fully hadronic mode to probe the $t\bar {t} H$ signal \cite{Buckley:2013lpa,Buckley:2013auc}, charged Higgs $tH^{\pm}$ signal \cite{Yang:2011jk} and top partner signal \cite{tprimehadron} have been done.


The early analysis on the search sensitivity of $b^\prime$ can be found in \cite{Skiba}, where a simple W-jet mass method was used. By using the semileptonic and dileptonic modes, the authors found that 1 TeV heavy quark can be reachable with 100 fb$^{-1}$ dataset. Another interesting work can be found in \cite{Holdom:2007ap}, where mainly using leptonic modes, Bob Holdom observed that it is difficult to use one cone to capture both boosted W boson and Top quark from $b^\prime$ decay. For the semileptonic mode studied in \cite{Holdom:2010fr,Holdom:2011uv}, it is observed that once all physics objects (say two top quarks, two W bosons, two $b^\prime$ quarks) can be reconstructed it is possible to extract the most crucial variables (like the mass bump of $b^\prime$) to suppress the SM background to a controllable level. A thorough study for top-partner can be found in \cite{AguilarSaavedra:2009es}, where leptonic and $b$ jet modes have been comprehensively analyzed. But the full hadronic mode has been left undone.

To our understanding, the study of hadronic mode of $t^\prime$ and $b^\prime$ has been untouched due to two major difficulties: 1) the full hadronic mode of the signal has high multiplicity of jets, and the combinatorics to find the characteristic parameters, like the masses of $b^\prime$ and $t^\prime$ are challenging; 2) Without characteristic variables for signal, it is difficult to distinguish signal from the SM multiple jet final states, say $t {\bar t} + \textrm{jets}$. In the study for charged Higgs boson \cite{Yang:2011jk}, two of our authors have noticed that the top tagger indeed can help to capture signal while maintaining suppression to the SM background even in full hadronic mode. It is observed that when the featured kinematic variables of signal are reconstructed, by using the multivariable analysis techniques, like the boost decision tree \cite{Roe:2004na,Yang:2005nz,Yang:2007pb} and neural network analysis, it is possible to pick out sufficient signal events when luminosity is large enough (say 100 fb$^{-1}$). Taking into account the recent quick development in tagging the boost objects \cite{Jetsubrev1,Boost2012}, it is well-motivated to explore the hadronic mode by adopting the recently developed hadronic top quark taggers and W boson taggers.

When an extra quark is heavy, massive objects in its decay final states like top quarks, W/Z/Higgs bosons, can be highly boosted.  Using the jet substructure techniques\cite{Jetsubrev1,Boost2012}, it offers a promising method to pick out possible signals while keep good suppression to the SM multiple jet final states. Based on Monte Carlo methods, these jet substructure techniques have been demonstrated to work quite well in searching for top partners and bottom partners \cite{HToppartnerJetsub,HEPTopTagger,Stoplep,StopJetsub,ToppartnerJetsub,StopSbottomJetsub}. Recently, theoretical understanding on jet physics from QCD side has brought new insights to the jet substructure and related phenomenological analysis methods, such as jet functions \cite{Berger:2003iw}, grooming\cite{Dasgupta:2013ihk}, quark/gluon separation \cite{Li:2011hy,Li:2012bw} etc. For a comprehensive review on the jet substructure based on first principle QCD calculation and monte carlo tools, we refer to Ref. \cite{Boost2012}.

In this work we use the hadronic top quark taggers and W boson taggers to explore the sensitivity of the LHC to $b^\prime$. We consider the process for $pp \to b^\prime {\bar b}^\prime$ at LHC by assuming that $b^\prime$ decays to top quark and W boson $100\%$ and study the heavy quark $b^\prime$ in the mass range $0.8$ TeV $<  m_{b^\prime} <1.5$ TeV. We propose a reconstruction procedure and demonstrate how to reconstruct all physics objects in the signal events. We further use multivariable analysis methods to optimize cuts and explore the sensitivity of the LHC to $b^{\prime}$.

This paper is organized as follows. In Sec. II, we briefly review the boosted massive object taggers which will be used in this work. In Sec. III, we present our phenomenological analysis. Finally, we make a discussion and give our conclusions in Sec. IV.


\section{Brief introduction to top taggers and W taggers}
At the LHC, there are large samples of W and Z bosons, Higgs, and top quarks with a transverse momentum $P_t$ that considerably exceeds their rest mass. In this kinematic regime, conventional reconstruction algorithms that rely on a one-to-one jet-to-parton assignment are often inappropriate, in particular for hadronic decays of such boosted objects. The technique of jet substructure has been developed to tag the boosted electroweak massive particles with hadronic decay \cite{Jetsubrev1,Boost2012,BDRS}. Roughly speaking, these tagging algorithms work in two steps: firstly cluster jets with a much larger radius parameter to capture the energy of the complete hadronic decay in a single jet; secondly use delicate discriminating variables to anatomize the internal structure of these fat jets in order to separate boosted objects from the large QCD background. Below we summarize several most common used algorithms for boosted top and boosted W on the market.

\subsection{Top-taggers}
Top quarks play an important role in understanding electroweak symmetry breaking and searching for new physics. Unlike the case at Tevatron where most of top quarks are produced near the threshold, at the LHC many boosted top quarks can be produced.
So, the jet substructure technique used to identify the boosted top from its hadronic decay has been developed in recent years.
For a recent review on top taggers, we refer to Ref.\cite{TopTaggerreview}.
Here, we give a brief review on John Hopkins top-tagger \cite{JHTopTagger} and
the HEPTopTagger (Heidelberg-Eugene-Paris) \cite{HEPTopTagger}.

{\bf A1. Johns Hopkins top-tagger }

After the success of BDRS algorithm \cite{BDRS} in Higgs search, Johns Hopkins group extended the BDRS algorithm to top study and proposed a top tagging method for the highly boosted top in hadronic decay \cite{JHTopTagger}. We will dub it as "JHTopTagger" for simplicity and use it in later analysis. As well-known, a boosted top from its hadronic decay looks like a fat jet with three hard cores.
Similar to the BDRS jet substructure method for boosted Higgs, the JHTopTagger firstly uses a large cone to cluster the
event in order to capture all the decay products and relevant radiations of a boosted top and then de-clusters the top jet to find three subjets inside the massive mother jet.

To resolve a fat jet into the relevant hard substructure from top decay, the following recursive procedure are applied in JHTopTagger\cite{JHTopTagger}:
 \begin{enumerate}
\item Four momenta of all particles from top decay are clustered into a massive jet with a large cone size $R$ ( CA algorithm are used in original paper),
\item Undo the last combination to get two objects $j_1$ and $j_2$. If the $P_t$ ratio of the softer jet $j_2$ over the original jet $j$ is too small, i.e., $P_{t_{j_2}}/P_{t_j}< \delta _p$ , throw the softer $j_2$ and go on to decluster on the harder one.
\item The declustering step is repeated until two separated hard objects are found. If any criterion below are satisfied, the declustering is failed: 1) both objects are softer than $\delta _p$ (2) two objects are too close, $ \Delta \eta + \Delta \phi < \delta_r $
 (3) the original jet is considered irreducible.
\item Declustering repeatedly on these two subjets will result in 2,3, or 4 hard objects.
\item It is required that the total mass of these subjets (only 3 or 4 hard subjets are considered) should be near $ m_t$ and the mass of two subjets among those resolved subjets should be in the $ m_W$ window. Furthermore, $W$ helicity angle $\theta_t$ should be consistent with a top decay due to the left handedness of the SM. Here, the helicity angle $\theta_t$ is defined in the rest frame of the reconstructed $W$ and is equal to the angle between the reconstructed top's fly-in direction and the fly-out direction of one of the two jets of $W$ decay products. Typically, the softer subjet in the lab frame are chosen to set the angle.
\end{enumerate}
The parameters involved in the method can be optimized event by event \cite{JHTopTagger}. In our study, for the LHC context, these parameters are fixed as below:
\bea
\label{eq:para}
\delta_p=0.19 \qquad \delta_r=0.1 \qquad |\theta_t|<0.65\,\,. \\ \nonumber
\eea

{\bf A2. HEPTopTagger}

Similar to the Johns Hopkins tagger, the HEPTopTagger~\cite{HEPTopTagger} (Heidelberg-Eugene-Paris) is developed to capture moderately boosted top and is firstly used to improve $t\bar{t}H$ searches. It begins with a large $R=1.5$ to cluster a fat CA jet. Such a large $R$ allows us to access top quarks down to a lower $P_t \sim 200$ GeV at the price of a large combinatorics of subjets and serious pile-up problem. The HEPTopTagger uncluster the fat jet using an iterative mass-drop criterion. At the meantime, it employs a filtering \cite{BDRS} procedure to pick up three hard subjets as candidates of top daughter jets and then test them with top kinematics.
The detailed delicate steps in HEPTopTagger to capture boosted tops are given:
\begin{enumerate}
\item The last cluster of the fat jet $j$ is undone to get $j_1$ and $j_2$. And then the mass drop
  criterion $\min m_{j_i} < \delta_m \cdot m_j$ determines if we keep $j_1$
  and $j_2$. A subjet with a large jet mass $m_{j_i} > 30~\gev$ are further decomposed; otherwise, the subjet is put to the list of relevant substructure.
\item The algorithm further uses the filtering procedure to construct one three-subjet combination with a jet mass closest to $m_t$ as the top candidates.
\item If the three invariant masses
  $(m_{12}, m_{13},m_{23})$ for the $P_t$ ordering subjets ${j_1, j_2, j_3}$ satisfy one of the following three
  criteria, accept them as a top candidate:
\begin{alignat}{5}
&0.2 <\arctan \frac{m_{13}}{m_{12}} < 1.3
\qquad \text{and} \quad
R_{\min}< \frac{m_{23}}{m_{123}} < R_{\max}
\notag \\
&R_{\min}^2 \left(1+\left(\frac{m_{13}}{m_{12}}\right)^2 \right)
< 1-\left(\frac{m_{23}}{m_{123}} \right)^2
< R_{\max}^2 \left(1+\left(\frac{m_{13}}{m_{12}}\right)^2 \right)
\quad \text{and} \quad
\frac{m_{23}}{m_{123}} > R_\text{soft}
\notag \\
&R_{\min}^2\left(1+\left(\frac{m_{12}}{m_{13}}\right)^2 \right)
< 1-\left(\frac{m_{23}}{m_{123}} \right)^2
< R_{\max}^2\left(1+\left(\frac{m_{12}}{m_{13}}\right)^2 \right)
\quad \text{and} \quad
\frac{m_{23}}{m_{123}}> R_\text{soft}
\label{eq:heptop}
\end{alignat}
\item For consistency, require the combined $P_t$ of the three subjets
  to be above 200~GeV.
\end{enumerate}
Here, the mass drop parameter $\delta_m$ and the mass windows parameters $R_{\min}$ and $R_{\max}$ are taken as $\delta_m=0.8$,
$R_{\min}=85\% \times m_W/m_t$ and $R_{\max}=115\% \times m_W/m_t$. The soft cutoff $R_\text{soft} =
0.35$ is supposed to remove QCD and W+jets background events.
The HEPTopTagger has an identification efficiency of roughly 40\% for top quarks with $P_t > 400$ GeV \cite{HEPTopTagger}.

We have compared the performance of these two top taggers and noticed that the HEPTopTagger can have a relative better performance, which can be attributed to the following two reasons: 1) the HEPTopTagger can capture not only highly boosted top but also intermediate boosted top; 2) due to more variables are used, the tagger can maintain a remarkable rejection to the SM background events even for the highly boosted top. Therefore, in the following study, we will adopt the HEPTopTagger to tag boosted top quarks in the signal events.

\subsection{W-taggers}

There are large samples of highly boosted electroweak massive particles, such as W bosons at the LHC.
The hadronic decay products of these massive particles will be collimated to form fat jets. These W-jets are different from QCD jets in two main aspects. Firstly, a W jet contains two hard subjets in similar energy and mass,
originated from the two quarks in the W decay, while a QCD jet usually has only one
hard subjet and asymmetric energy-flow distribution. On the other hand, a QCD jet is initiated from a color triplet or octet, which is color-connected to the beam or the other side of the event. Whereas, the two subjets of a W-jet are from a color singlet and they tend to correlate to each other in color. Along these lines, many sophisticated W-tagging tools has been developed to pick out the highly boosted W bosons from backgrounds \cite{Jetsubrev1,CMSsearchWjet,CMSWTagger,TMVAWTagger}.

{\bf B1. CMS W tagger }

The LHC experiment has employed the W tagging algorithms to search for new physics \cite{CMSsearchWjet,CMSWTagger}. Here, we briefly introduce the W tagging algorithm used by CMS collaboration (CMSWTagger) \cite{CMSWTagger}, which mainly use pruning \cite{pruning} and mass drop \cite{BDRS} methods. The algorithm can be applied to a massive jet and is given as follows.
\begin{enumerate}
\item The pruning method: the clustering history for a fat and massive jet is checked at every step. For a merging step say $(i+j \rightarrow p)$, two conditions are examined:
\beq
z_{ij} \equiv \frac{\min(P_{ti}, P_{tj} )}{P_{tp}}> z_{\textrm{cut}}
\eeq
and
\beq
R_{ij} < D_{\textrm{cut}} = 2 \times \frac{m_J}{{P_t}_J}
\eeq
If this $(i + j\rightarrow p)$ step does not satisfy these two conditions, $i$ and $j$ will not be merged and instead
the softer of the two clusters is removed.
\item The mass drop method: the total mass of the above pruned jets are required to be in the W mass window $70 \gev < m_{jet} < 100 \gev$. Undoing the last clustering iteration of the pruned jet to get two subjets. The ratio of masses of the hardest subjet ($m_1$) and the total pruned jet mass is defined as the mass drop
$\mu= \frac{m_1}{m_{\textrm{jet}}}$. To discriminate against QCD jets, the mass drop is required to satisfy
$\mu < 0.4$.
\end{enumerate}

{\bf B2. Multivariate analysis W tagger }

Unlike the case in top tagger, there are only few orthogonal variables in the two body decay of a highly boosted W.
In Ref \cite{TMVAWTagger}, a jet substructure algorithm with multivariate analysis was proposed for distinguishing highly boosted hadronically
decaying $W$ from QCD jets. The algorithm, dubbed it as "TMVAWTagger", selects 25 most useful variables and combines them using the Boosted
Decision Trees method. These variables include the masses and $P_t$'s after jet grooming, planar flows, $P_t$ $R$-cores, etc.
The detailed steps of the TMVAWTagger are presented as below:
\begin{enumerate}
\item Begin with fat jets with large $R=1.2$ and then use the filtering/mass drop \cite{BDRS} to identify W jet candidates.
\item Apply a filtering step to get the leading three filtered subjets. The jet mass of combination of these jets is required to be within the
mass window (60,100) GeV.
\item After the mass window cut, the original unfiltered fat jets are treated by using the multivariate analysis to maximize the efficiency.
The 25 variables used in the analysis are:
\beq
m_{\textrm{jet}},\ c_{P_t}(0.2-1.1),\ \textrm{sens}^{m,{P_t}}_{\textrm{filt,trim,prun}},\ P_f,\ P_f(0.4),\ \frac{{P_t}^{\textrm{sub1,sub2}}}{P_t},\  \frac{m^{\textrm{sub1,sub2}}}{m},\ \Delta R_{\textrm{sub}}, n_{\textrm{sub}}.
\eeq
Here, $c_{P_t}$'s are $P_t$ $R$-cores \cite{TMVAWTagger}, from 0.2 to 1.1 by 0.1 and $\textrm{sens}^{m,{P_t}}_{\textrm{filt,trim,prun}}$ represent 6 grooming
\cite{BDRS,pruning,trimming} sensitivities. $P_f$ and $P_f(0.4)$ are the planar flow parameters for the original jet and for the highest
$P_t$ subjet from reculstering with $R=0.4$, respectively. $\Delta R_{\textrm{sub}}$ is the distance between the two leading subjets and $n_{\textrm{sub}}$ is the total number of subjets after the filtering process.
\end{enumerate}
By using the multivariate analysis, the TMVAWTagger can be quite robust to reject background. Nonetheless, to simplify the current study, we use the CMSWTagger and choose a smaller cone to capture the W bosons from the decay of $b^\prime$.

\section{Numerical Results and Analysis}
At the LHC, $b^\prime$ pair is mainly produced by the gluon fusion process. The cross section of $p p \to b^\prime {\bar b}^\prime$ has been studied at reference \cite{Cacciari:2008zb} up to NLO + NLL. For the collision energy $\sqrt{s}=14$ TeV, when $m_{b^\prime}$ is around 1 (2) TeV, the cross section can be 69 (0.3) fb. When we assume that $b^\prime$ decays 100$\%$ to top and W boson, its decay width is less than $10\%$, so the narrow width approximation is still held.

\subsection{Features of signal}
In order to determine the right parameters to tag both top quarks and W bosons, we analyze the distributions of cone sizes of top quarks and W bosons in the final states at the parton level. We call the W bosons directly from the $b^\prime$ decay as the isolated W bosons and label them as $W^{\textrm{iso}}$. In contrast, we call the W bosons from the top quark decays as non-isolated W bosons and label them as $W^{\textrm{non}}$.

\begin{figure}[!htb]
\begin{center}
\includegraphics[width=1.0 \columnwidth]{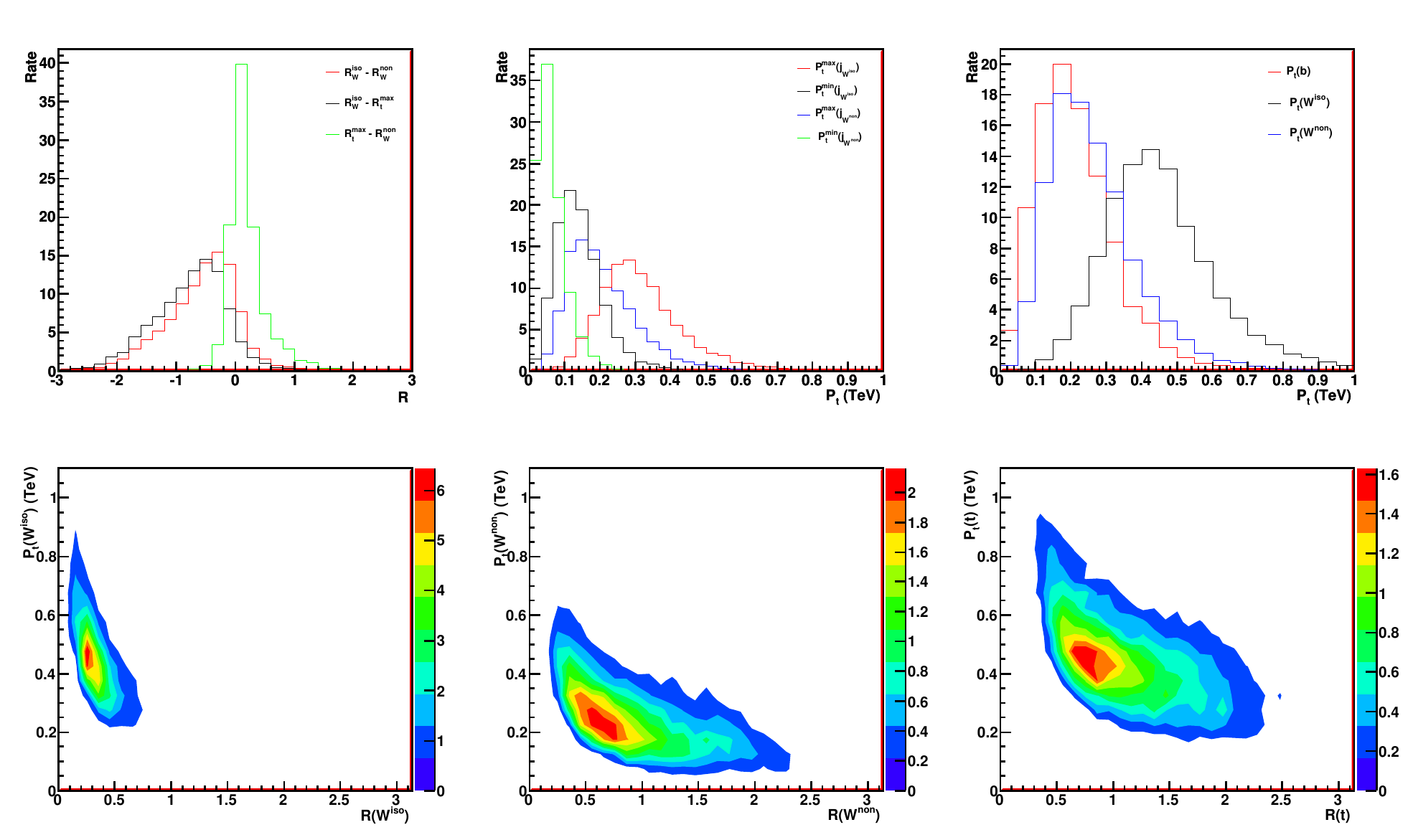}
\caption{The momenta distributions and angle separations of jets from top quarks and W boson decay are demonstrated at parton level with $m_{b^\prime}=1$ TeV as a show case.
\label{rboost}}
\end{center}
\end{figure}

The correlations between the transverse momenta and the largest angle separation between two jets of two types of W bosons are shown in Fig. \ref{rboost}. In order to estimate the value of cone parameter $R$ to cluster two jets from W boson hadronic decay into one fat jets, we define $R^{\textrm{iso}}_W$ as $R^{\textrm{iso}}_W=\textrm{max}(R(P(W), P(j_1)),R(P(W), P(j_2)))$, where $P(j_1)$ and $P(j_2)$ label the momenta of two daughter jets from the $W$ boson decay and $P(W)$ labels the momentum of the $W$ boson. From the plots, it is obvious that these two types of W bosons can be distinguished from their transverse momenta and angle separations. The most probable value of angle separation between two jets from the isolated W bosons is around $0.3$ and is smaller than that from the non-isolated W bosons, which is around $0.6$. The most probable transverse momenta of isolated W boson is around 500 GeV. It is larger than that of the non-isolated W bosons, which is around 300 GeV. These kinematic features can be utilized to determine the correct combinations of jets.

We also evaluate the largest angle separation of three partons from the top quark decay, which is defined as $R_t^{\textrm{max}} = \textrm{max}(R(P(t),P(j_1)),R(P(t),P(j_2)),R(P(t),P(j_3)))$. The most probable value is around $0.8$, which tells us that in order to capture the boosted top quarks, we'd better use a cone size parameter around $1.1$ or so. By comparing the middle and right plots in the lower row given in Fig. \ref{rboost}, we can conclude that, at most of time, the $R(t)$ is determined by the angle separation of non-isolated W boson, which can also be read out from the left plot in the upper row from the curve $R^{\textrm{max}}_t - R^{\textrm{non}}_W$.

\begin{figure}[!htb]
\begin{center}
\includegraphics[width=1.0 \columnwidth]{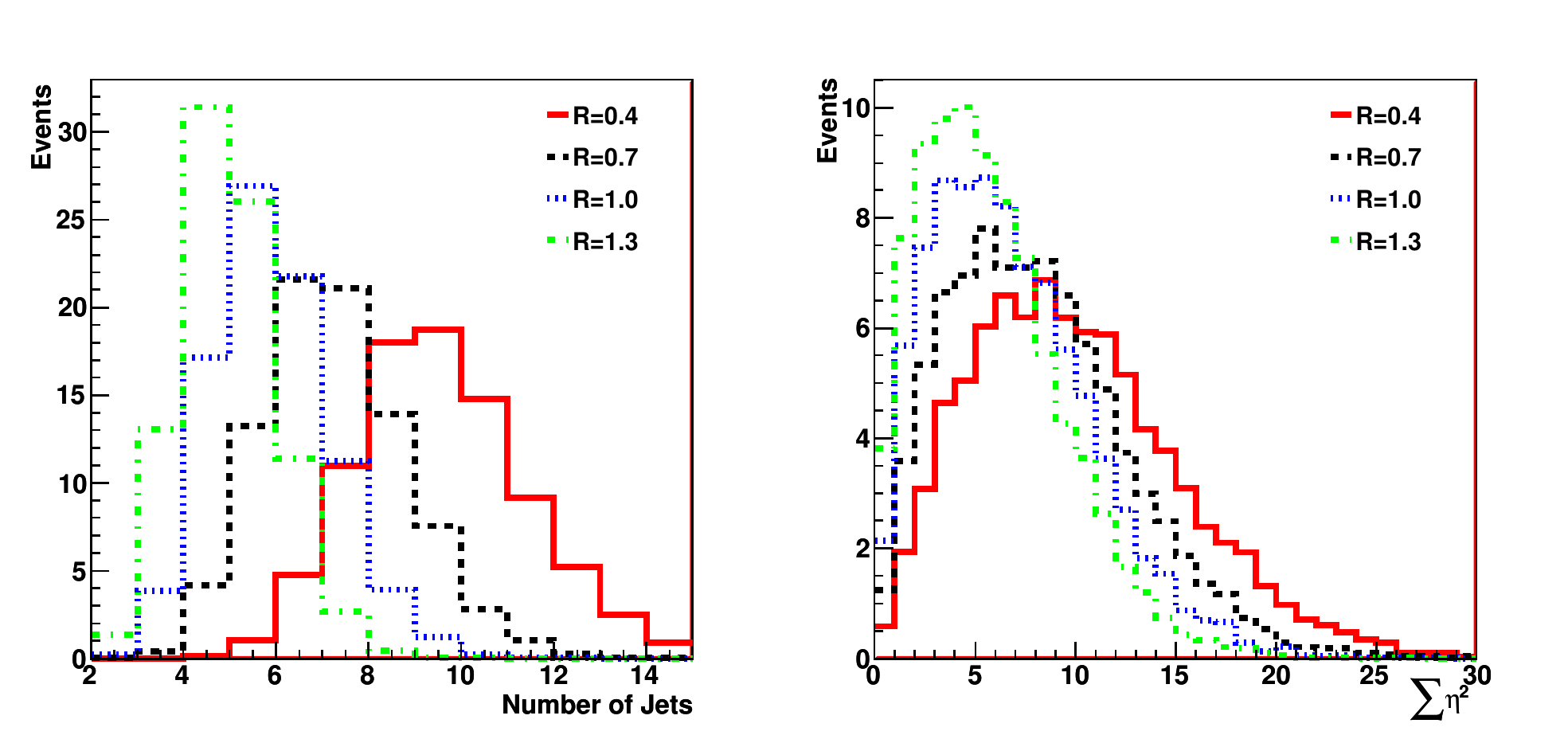}
\caption{The number of jets and the observable $\sum\eta^2(j)$ changed with the cone parameter $R$ in the anti-Kt jet algorithm (where detector effects are simulated by PGS) with $m_{b^\prime}=1$ TeV as our show case.
\label{nj-eta2}}
\end{center}
\end{figure}

We also examine the number of jets and one observable defined as $\sum\eta^2(j)$ with the change of the cone size parameters in the anti-kt algorithm, as demonstrated by Fig. \ref{nj-eta2}. It is obvious that when the size of cone parameters for jet algorithm is changed, the number of jets can be changed, so as some kinematic observable, like $\sum\eta^2(j)$. There are also some observables, like the centrality, which are found to be insensitive to the change of cone
parameter.

\begin{figure}[!htb]
\begin{center}
\includegraphics[width=1.0 \columnwidth]{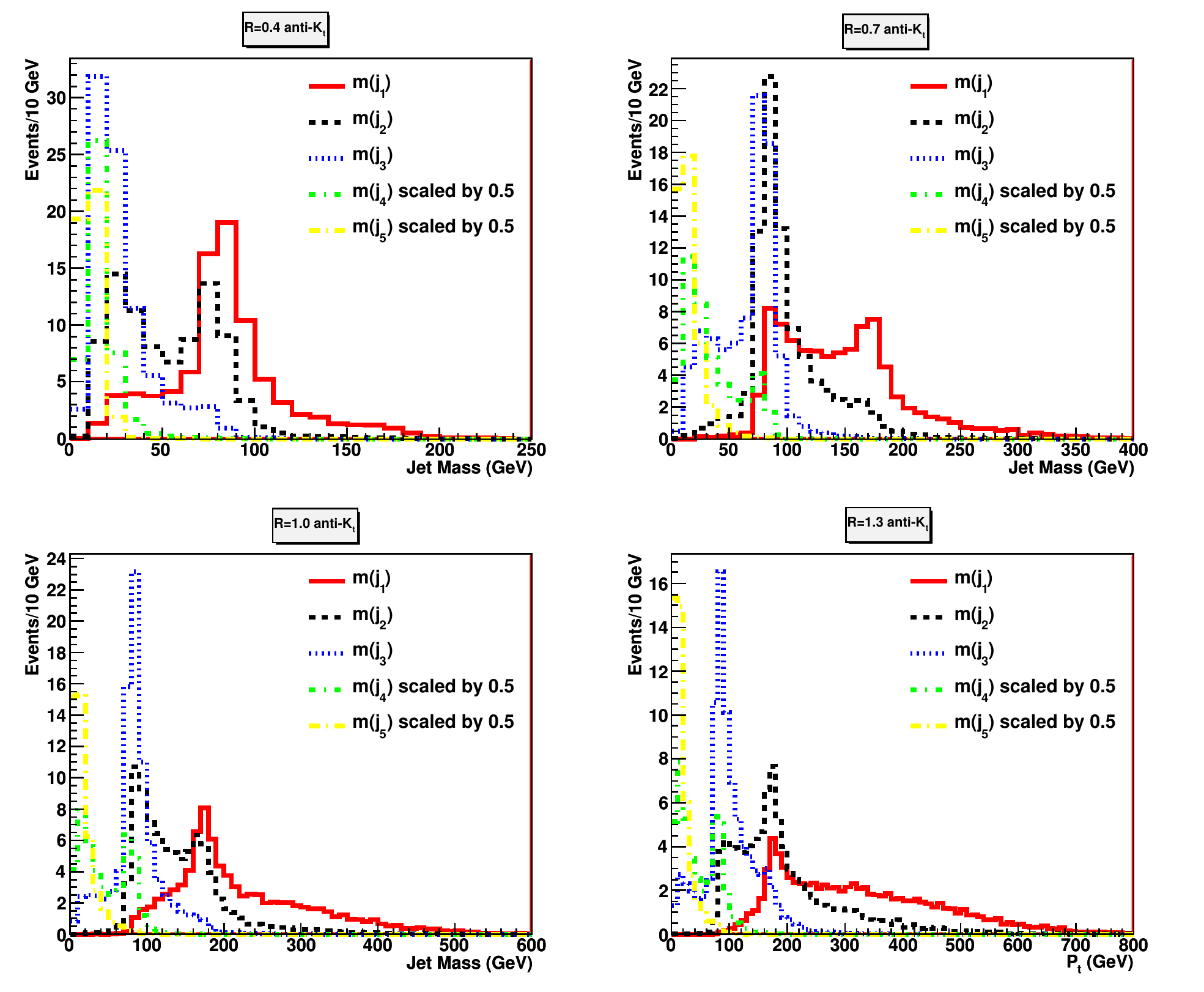}
\caption{The jet mass distribution for the leading 5 massive jets in each event is shown, where jets are clustered by the anti-Kt jet algorithm (where detector effects are simulated by PGS) with $m_{b^\prime}=1$ TeV.
\label{jetmass}}
\end{center}
\end{figure}

We also examine the jet mass distribution with the change of cone parameter $R$ in the anti-Kt jet algorithm, as shown in Fig. \ref{jetmass}. We would like to mark a few salient features from Fig. \ref{jetmass}. \begin{itemize}
\item When $R=0.4$, it is found that more than $80\%$ of the first leading jet can has a jet mass in the W-mass window (the window is defined as $|m(j) - m_W^{\textrm{PDG}}| < 20$), while more than $50\%$ of the second leading jet can have a jet mass in the W-mass window. Most of these massive jets are from the isolated W bosons, which is consistent with our analysis at the parton level shown in Fig. \ref{rboost}.
\item When the cone parameter $R$ is changed to $0.7$, more than $40\%$ of first leading massive jet has a jet mass in the top quark mass window (the window is defined as $|m(j) - m_t^{\textrm{PDG}}| < 30$) and $30\%$ of the first leading massive jet has a mass in the W-mass window. It is remarkable that more than $80\%$ of the second and third leading massive jets are in the W-mass window. This indicates that the optimized cone parameters for W-jets should be around $R=0.7$.
\item When the cone parameter $R$ is changed to $1.0$, more than $50\%$ of the first leading massive jet has a jet mass in the top quark mass window. There are around $25\%$ of the second leading massive jet in the top quark mass window and $60\%$ in the W-mass window. More than $85\%$ of third leading massive jet is in the W-mass window. It is remarkable that there are around $40\%$ the fourth massive jet in the W-mass window.
\item When the cone parameter $R$ is changed to $1.3$, more massive jets can be in the top quark mass window. Compared with the plot for $R=1.0$, we can read out that the optimized cone parameter $R$ for a fat jet as a top quark should be around $1.0<R< 1.4$ or so. Nevertheless, the optimized cone parameter should also take into account the behavior of background events. Another  noticeable point is that the mass of the fifth jet indicates the mass dependence on the $R$.
 \end{itemize}

Although some massive jets are outside the W-mass window or the top quark mass window, it is expected that the jet tagging techniques should help us to find their identities.

It is obvious that different $R$'s can reveal parts of the full information of signal, a better way to separate signal and background is to utilize all available information with different $R$'s when the computing time is allowed. When the computing resources are limited, we can use some tight preselection rules to choose the most relevant events in our analysis.

Obviously, when the mass of $b^\prime$ increases from 1 TeV to 2 TeV, both W bosons and top quark become more energetic and their decay products can be more collimated. We observe that smaller cone size parameters can capture a considerable fraction of W bosons and top quarks, respectively, which is understandable from rule of thumb $R=2 \,m/P_t$.

It is remarkable that the origin of jet masses of top quarks and W bosons is different from that of QCD jet. The masses of those fat jets from top quark and W boson are from EW symmetry breaking, while the mass of QCD jet is from the collinear and infrared which can lead to wrong combination of pseudo-jets into a massive QCD jet.

\begin{table}[th]
\begin{center}%
\begin{tabular}
[c]{|c|c|c|c|c|c|c|c|c|}\hline
      & \small (0.5,1.3)  & \small (0.6,1.2)  & \small (0.6,1.3) & \small (0.6,1.4)   & \small (0.7,1.2) & \small (0.7,1.3)  &\small (0.7,1.4) &\small (0.8,1.4)   \\ \hline
(0,0) &  1 & 0   &  0  &0    & 1 & 0  & 0    & 1  \\ \hline
(0,1) &  4 & 4   &  3  &3    & 4 & 4  & 3    &  4 \\ \hline
(1,0) &  4 & 3   &  3  &3    & 3 & 3  & 3    & 4  \\ \hline \hline
(0,2) &  5 & 6   & 5   &4    & 6 & 5  & 4    &  4\\ \hline
(0,3) &  1 & 1   & 1   &1    & 1 & 1  & 1    &  0\\ \hline
(1,1) &15 & 14 & 14 &13  &14& 13 &13&   13 \\ \hline
(1,2) &11 & 14 & 13 &12  &14& 13 &12&  11 \\ \hline
(1,3) &  1 & 1   & 1   &1    & 1 & 1   & 1 &  1 \\ \hline
(2,0) &  6 & 5   & 5   &6    & 5 & 6   &  6&  7  \\ \hline
(2,1) &10 & 9   & 10 &10  & 9 & 9   &10&  9  \\ \hline
(2,2) &  3 &  4  & 4   &3    & 4 & 4   & 3 &   3 \\ \hline
\end{tabular}
\end{center}
\caption{The ($R_W$-$R_t$)-dependence of the fractions tagged hadronic top quark and tagged W boson in the process
$pp \rightarrow b'\bar{b}'  \rightarrow tW^-\bar{t}W^+$ are examined with $m_{b'}=1$ TeV as a show case. At the head of each row, the first digit means the number of tagged top quark, and the second digit means the number of tagged W boson.
For example, (1,2) means one tagged top quark and two tagged W bosons. For both top quark and W boson taggers, we adopt the CA jet algorithm.
We use HEP top tagger for $R=1.0-1.4$ to find tagged top quarks. We adopt the CMS W tagger to find tagged W bosons. Numbers in the table denote
percentage. Numbers less than 1 are omitted. }%
\label{tagmodes}%
\end{table}

In our Monte Carlo study, the signal events are generated by Madgraph/MadEvent \cite{Madgraph} and background events by Alpgen \cite{Alpgen}. We have used the MLM matching \cite{MLM} to avoid the double counting issue. These events are fed to DECAY to generate full hadronic decay final states and pass to PYTHIA \cite{Pythia} to simulate showering, fragmentation/hadronization, initial state radiation, final state radiation, and multi-interaction as well. After that, fastjet \cite{Fastjet} and SpartyJet\cite{Spartyjet} are used to perform jet clustering and massive object tagging analysis.

We would like to make a comment on the modeling of the SM background events. It is highly nontrivial to model the SM background events with 10 jets. For example, we noticed that the QCD jet sample is quite difficult to be generated by the MC tools on the market. Instead, by choosing the jet parameter $P_{t_{\textrm{min}}}>100$ GeV in the Alpgen, we generate exclusive datasets of $ 2 j$, $3 j$, $4 j$, $5 j$ events, and an inclusive $6 j$ data sample. In our later analysis, we demand $n_j > 7$ and the $P_t$ of the leading two jets momenta is larger than 200 GeV, so those extra jets can only be produced from the initial state radiation and finial state radiation. In this sense, for QCD jets data sample, our treatment can be regarded as leading order approximation. It is also true for case with the background events $W/Z+$ jets and diboson jets, where at most, Alpgen can allow us to generate $W/Z+6$ jets at matrix element level. Nonetheless, we noticed that these two types of background can be efficiently suppressed by $b$ taggings and the kinematics cuts introduced below. While for the $t{\bar t}$ dataset, we have used Alpgen to generate exclusive datasets $t{\bar t}+ 0j$, $t{\bar t}+1j$,$t{\bar t}+2j$, $t{\bar t}+3j$, and an inclusive dataset for $t{\bar t}+4j$, and we merge these exclusive and inclusive datasets into an inclusive data for $t{\bar t}$ type background. For $t{\bar t}$ type background events, at most, we can marginally generate final states with 10 parton by the matrix elements. Keeping this fact in mind, the treatment to the SM background demonstrated in this work can only serve as a leading order approximation.

\subsection{A proposed reconstruction procedure}
To extract useful information of signals, we propose to cluster jets with three different sizes and the following reconstruction procedure to find all objects of an event:
\begin{itemize}
\item[1)] For the small size jets, we use the anti-K$_t$ algorithm with jet parameter $R=0.4$. We only consider high jet multiplicity events with $n_j \geq 9$ and $H_t > 1.5 m_{b^\prime}$. When there are massive jets in the event, we demand the number of massive jets $n_{m}$  (defined as $m(j_i) > 60$) and the number of non-massive jets should satisfy $2 \,n_{m} + n_j \geq 10$.
\item[2)] For the mediate size jets, we use the CA algorithm with jet parameter $R=0.6$. These mediate jets are supposed to find massive objects, especially the massive W bosons. Some of highly boosted top jets can be also found.
\item[3)] For the largest size jets, we use the CA algorithm with jet parameter $R=1.3$, we find massive objects, especially the boosted top quarks. We label the number of tagged top quarks as $n_t$.
\item[4)] We identify non-isolated W bosons by using the massive jet found at step 2 and 3. We examine whether each of identified W bosons at step 2 is in the cone of the identified top quarks. If a tagged W boson is in the cone of a top jet, we label it as a non-isolated W boson. If not, it will be labelled as un-used and will be used to further determine the missing objects. The number of un-used tagged W bosons is denoted as $n_W$.
\item[5)] We identify isolated small jets with $R=0.4$ which is neither in the cone of W bosons nor in the cone of top quarks. And we use them to  reconstruct all missing objects, like top quark(s), W boson(s) and $b^\prime$s. To avoid the severe issue of wrong combinatorics, we throw away events when less than two objects are identified, i.e. $n_t + n_W < 2$.
\end{itemize}

There are several comments in order:

1) We use the minimum $\chi^2$ approach to find the missing objects. For example, if in one event, we have tagged two top quarks with $n_t=2$ and $n_W=0$, then the rest of work is to reconstruct two W bosons by using the rest of isolated small jets. The $\chi^2$ is defined as $\chi^2 = \frac{(m_{12} - m^{\textrm{PDG}}_W)^2}{\sigma_W^2} + \frac{(m_{34} - m^{\textrm{PDG}}_W)^2}{\sigma_W^2}$. If in one event, we have tagged two W bosons with $n_t=0$ and $n_W=2$, then the $\chi^2$ is constructed as
$\chi^2 =  \frac{(m_{12} - m^{\textrm{PDG}}_W)^2}{\sigma_W^2} + \frac{(m_{123} - m^{\textrm{PDG}}_t)^2}{\sigma_t^2} +
\frac{(m_{45} - m^{\textrm{PDG}}_W)^2}{\sigma_W^2} + \frac{(m_{456} - m^{\textrm{PDG}}_t)^2}{\sigma_t^2} $, where we have taken into account the possibility that any one of or both W bosons could come from top quark decays.

2) Once two top quarks and two W bosons have been reconstructed, we choose the value of $m^{\textrm{rec}}=m_{\textrm{min}}$ which can minimize the $\chi^2(m)=\frac{(m_{b_1^\prime} - m)^2}{(\sigma_b^\prime)^2} + \frac{(m_{b_2^\prime} - m)^2}{(\sigma_b^\prime)^2}$ as the reconstructed $b^{\prime}$ mass, where the mass of $b_i^\prime$ is the combination of a pair of top and W boson, where both two possible combinatorics have been taken into account.

From the results presented in Table \ref{tagmodes}, we observe that if we require that $n_t=2$ and $n_W=2$, there are only $3-4\%$ of signal can be taken into account. While with $n_t + n_W \geq 2$ and the proposed reconstruction procedure, more than 50$\%$ of signals can be reconstructed successfully, while less than $1 \% $ of the dominant background $t{\bar t} + nj$ can pass through this reconstruction. These events can be correctly triggered at LHC collaborations due to both large $H_t$ and large number of jets with the standard anti-K$_t$ jet algorithm with the cone parameter $R=0.4$. To further reduce reconstruction time, we choose $n_t + n_W \geq 2$ and $n_t \geq 1$ as our reconstruction conditions. It is found that by choosing this selection rule we can achieve similar results.

In Table \ref{tagmodes}, fractions of tagged objects in the process $pp \rightarrow b'\bar{b}'\rightarrow tW^-\bar{t}W^+$ with different $R$ are provided. This Table provides important hints as how to capture our signal. There are a few comments in order. 1) The optimization in different cone sizes for $R_W$ and $R_t$ can affect the signal in $10\%$ percentage level. Therefore, for different masses of ${b^\prime}$, it might be useful to optimize $R_W$ and $R_t$. 2) It is crucial to capture top quark and W boson jets in our reconstruction procedure. 3) When $R_W$ becomes too large (say larger than 0.8) , the fraction of signal events in the reconstruction procedure becomes fewer due to the failure of W-tagger.

\subsection{Signal and background discrimination}
\begin{table}[th]
\begin{center}%
\begin{tabular}
[c]{|c|c|c|c|}\hline
 & $b^\prime {\bar b}^\prime$ [fb] & $ t {\bar t} +\textrm{jets} $  [fb]  & $t {\bar t} +\textrm{W}+ \textrm{jets} $  [fb] \\ \hline \hline
$\sigma$ $\times $ branching fraction $\times$ btagging &  6.0  &  $1.67 \times 10^5$ & $55.6$ \\ \hline \hline
$H_T > 1500$ GeV\& ${\cal C} > 0.55$ \& $n_j>7$ &  4.2  & $2.3 \times 10^3$  & 2.5\\ \hline \hline
Reconstruction \& $m_{b^\prime} > 650$ GeV & 2.2 & 147.2 & 0.2 \\ \hline
MLP ($N>0.9$) & $1.5 $ & $3.7$ & 0.02 \\ \hline
\end{tabular}
\end{center}
\caption{The cut efficiencies for signal and main background processes are shown here, where the signal is for the case $m_{b^\prime}=1 \textrm{TeV}$ .
The cross section of $t{\bar t}$ is taken as 945 pb from \cite{Cacciari:2008zb}, we assume all heavy particle decay in the hadronic mode.
The optimized cut efficiencies for both MLP and BDT methods are provided for comparison. We cluster jets by using the standard anti-Kt algorithm with the cone parameter $R=0.4$. }%
\label{cuteff}%
\end{table}

In order to suppress the tremendous QCD background events and make hadronic mode doable, at the preselection level for further analysis, we demand that the scalar sum of transverse momentum of all final states must be larger than $\frac{3}{2} m_{b'}$ and the centrality of each event must be larger than $0.55$. Furthermore, we require that at least two hadronic heavy objects (i.e. $n_t + n_W \geq 2$ ) must be identified and two $b$ jets must be tagged. In this work, we assume that $b$ tagging efficiency as 0.6 with a rejection factor 300. We further demand that there must be more than 7 jets with $P_t(j)>20$ GeV and the transverse momentum of leading two jets should be larger than 200 GeV in each event when jets are clustered with jet parameter $R=0.4$. With these conditions, we observe that the background of QCD multiple jets are highly suppressed by the conditions of leading two jets $P_t(j_1)>300$ GeV and $P_t(j_2)>200$ GeV, $H_t$, $b$ taggings and jet numbers. Similarly, the background of $t {\bar t} + W$ + jets is also suppressed significantly by $H_t$ and jet numbers.  After these conditions for preselection, we find that the dominant background is $p p \to t {\bar t} + \textrm{jets}$.

We have considered the contributions of diboson (WW, ZZ, etc.) backgrounds, $t{\bar t} + Z  +$ jets, and have found that these types of reducible background events can be safely neglected after imposing preselection rules. Diboson background are also generated by Alpgen with the number of jets upto 3. Such type of background can also be heavily suppressed by the requirement of boosted objects and reconstruction criteria. Background events from $t{\bar t}+ Z +$ jets with the number of jets upto 3 are very similar to  $t{\bar t}+ W +$ jets, which can at most reach to 1 percent of the signal after all cuts, therefore we neglect it here.

For the irreducible background $t {\bar t} + WW +$ jets with the number of jets upto 2 by using Madgraph5, we noticed that the cross section (3 fb for hadronic mode) of this type of background is close that of our signal when $m_{b^\prime} = 1$ TeV, but the $H_t$ cut and the requirements in leading jets, the kinematics of reconstructed W bosons (two W bosons in our signal have very large $P_t$ and are highly boosted), and the mass bump of $b^\prime$ (the background events occur near the threshold region) can help to remove such type of background down to a percent level of the signal after all cuts. Therefore, we neglect such type of background in Table \ref{cuteff}. For the same reasons, we also neglect $t {\bar t} t {\bar t}$ background with the number of jets upto 4 by using Alpgen, of which the cross section is 2 fb for fully hadronic mode.

\begin{figure}[htbp]
 \hspace{0.0cm}
\includegraphics[width=6.5in]{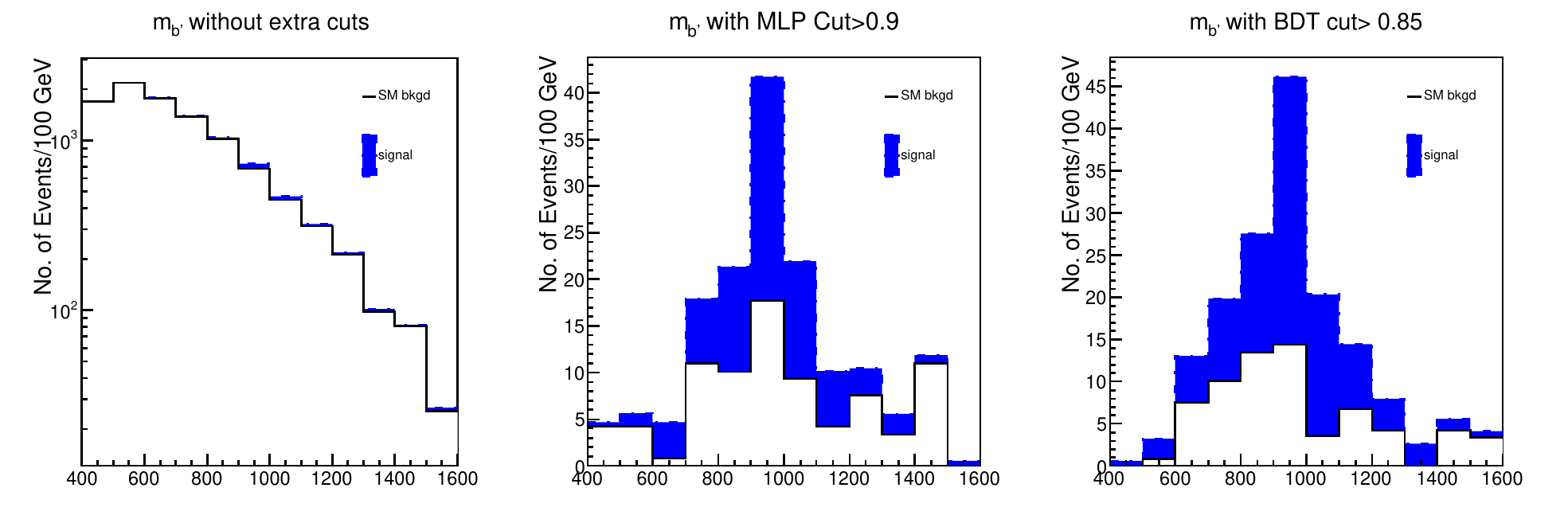}
\caption{The distributions of reconstructed mass bump of $b^\prime$ before/after  discrimination cuts are shown.
}
\label{figmass}
\end{figure}

In Table \ref{cuteff}, we illustrate how signal and background change with our preselection conditions. From it, it is observed that before reconstruction the dominant background is $p p \to t {\bar t} + \textrm{jets}$. As demonstrated in left plot of Fig. \ref{figmass}, even after the reconstruction, it is still quite challenging to find the signal, the total background is around two order of magnitude larger than our signal. In order to achieve a better $S/B$ and a better significance, obviously a dedicate signal background discrimination analysis is needed.

\subsection{Multivariate analysis}
Considering that there are 10 physics particles in our final state at parton level, the dimension of phase space of signal is 30 or so without taking into account the phase space of background events. As done in works \cite{Holdom:2010fr,Holdom:2011uv,Yang:2011jk}, in this work we have adopted two Multivariate Analysis methods: the neural network (multilayer perceptron) and the boosted decision tree.

\begin{figure}[!htb]
\begin{center}
\includegraphics[width=0.45 \columnwidth]{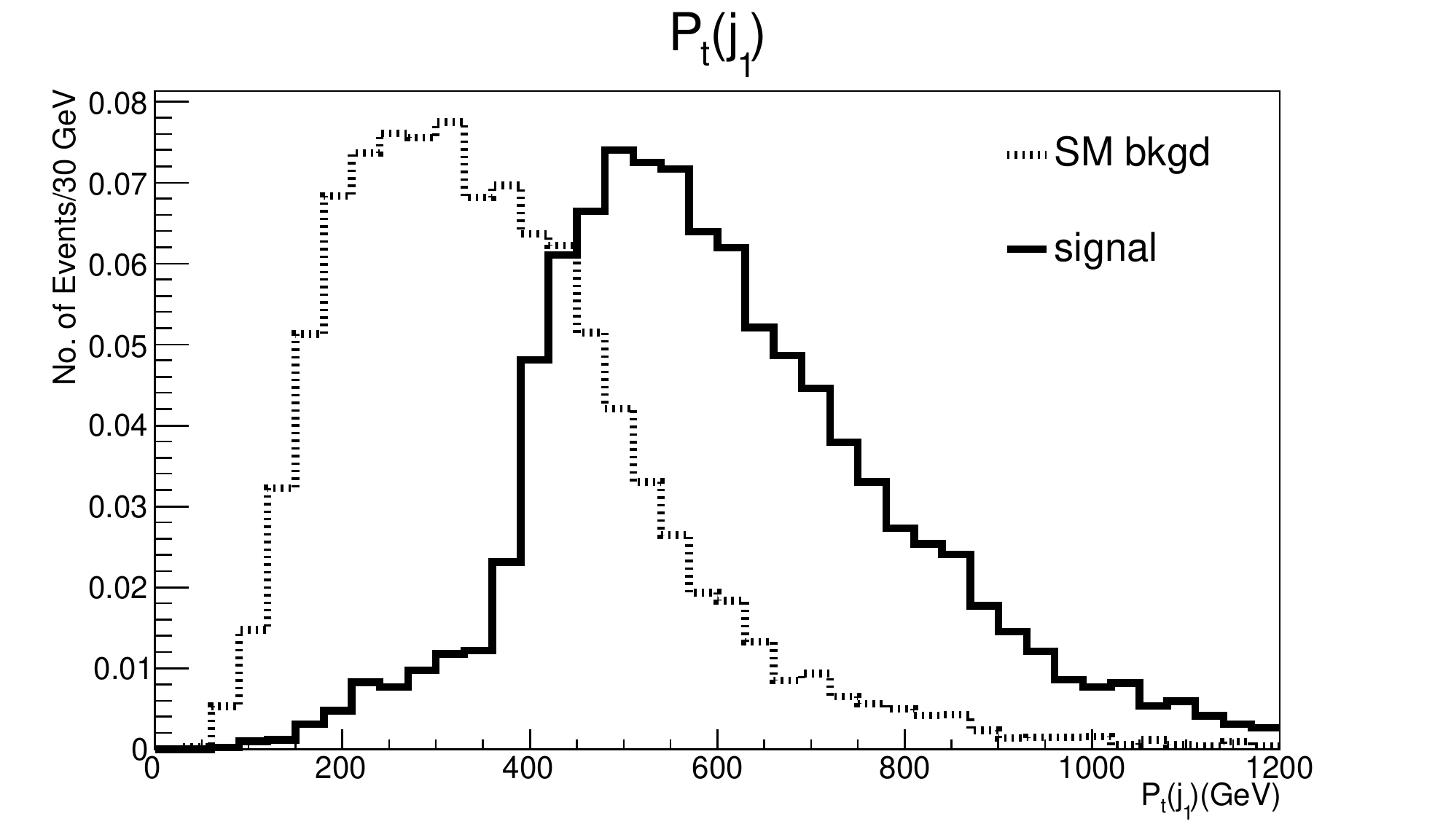}  \includegraphics[width=0.45 \columnwidth]{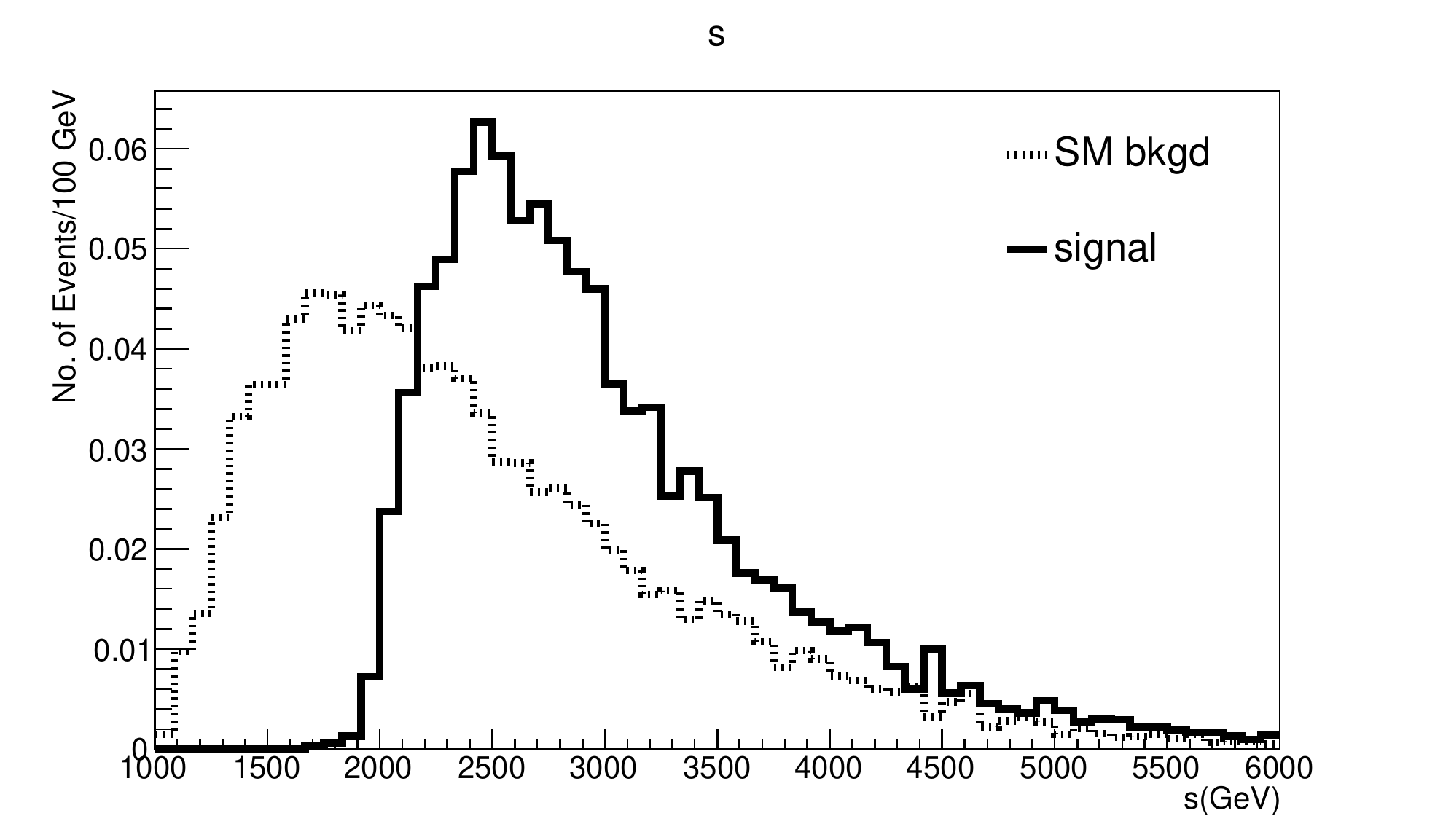} \\
\includegraphics[width=0.45 \columnwidth]{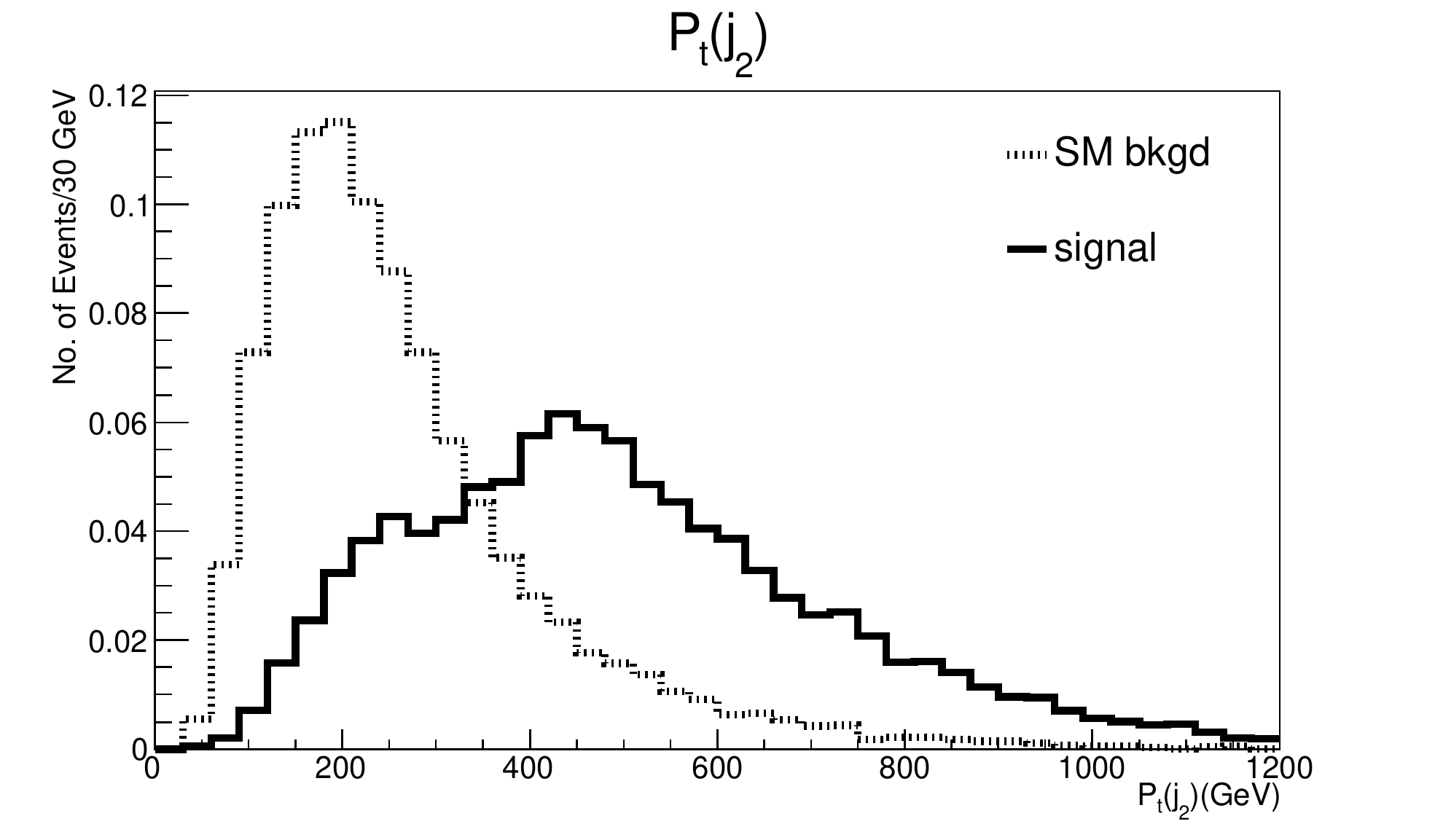}  \includegraphics[width=0.45 \columnwidth]{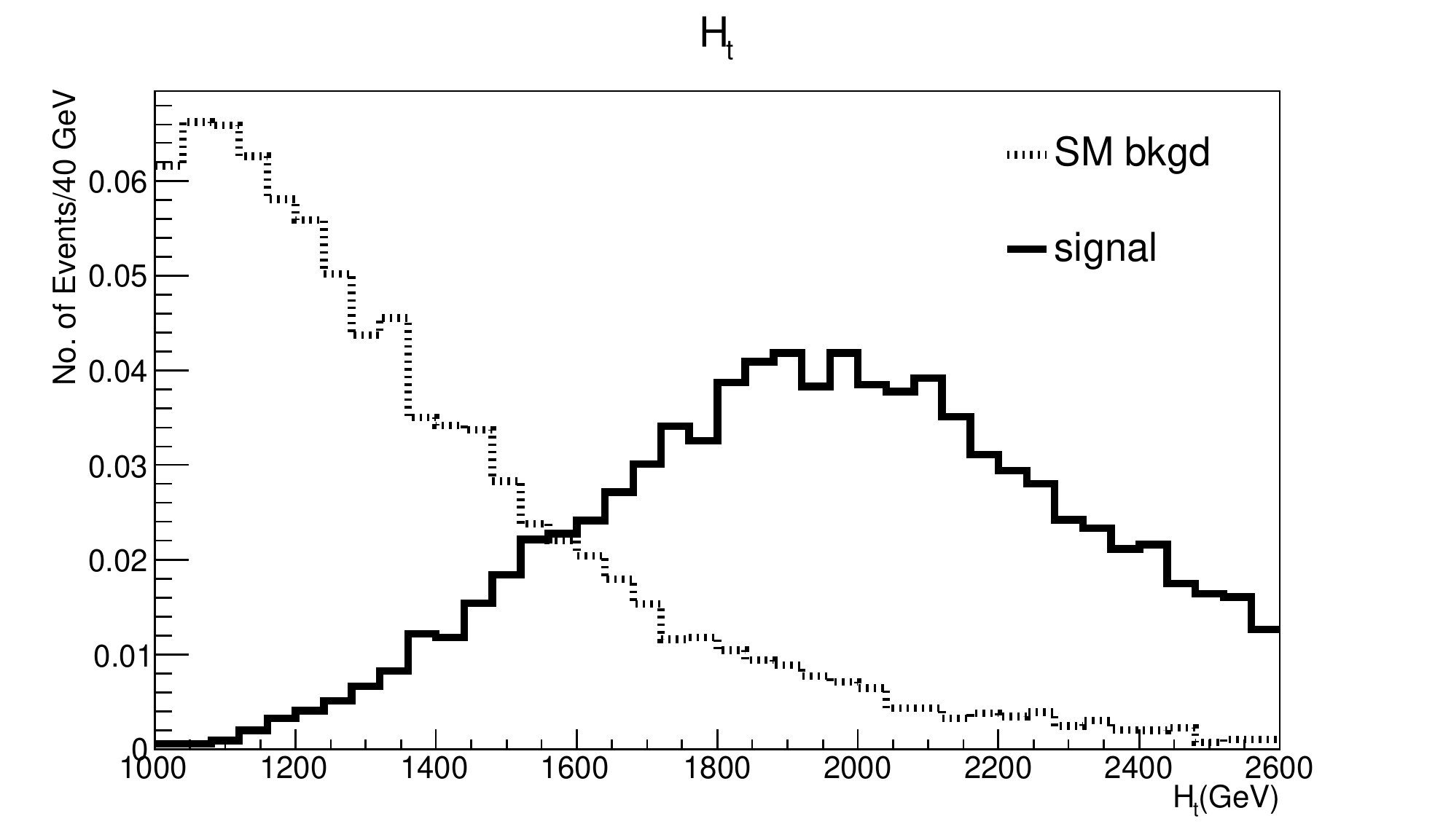}   \\
\includegraphics[width=0.45 \columnwidth]{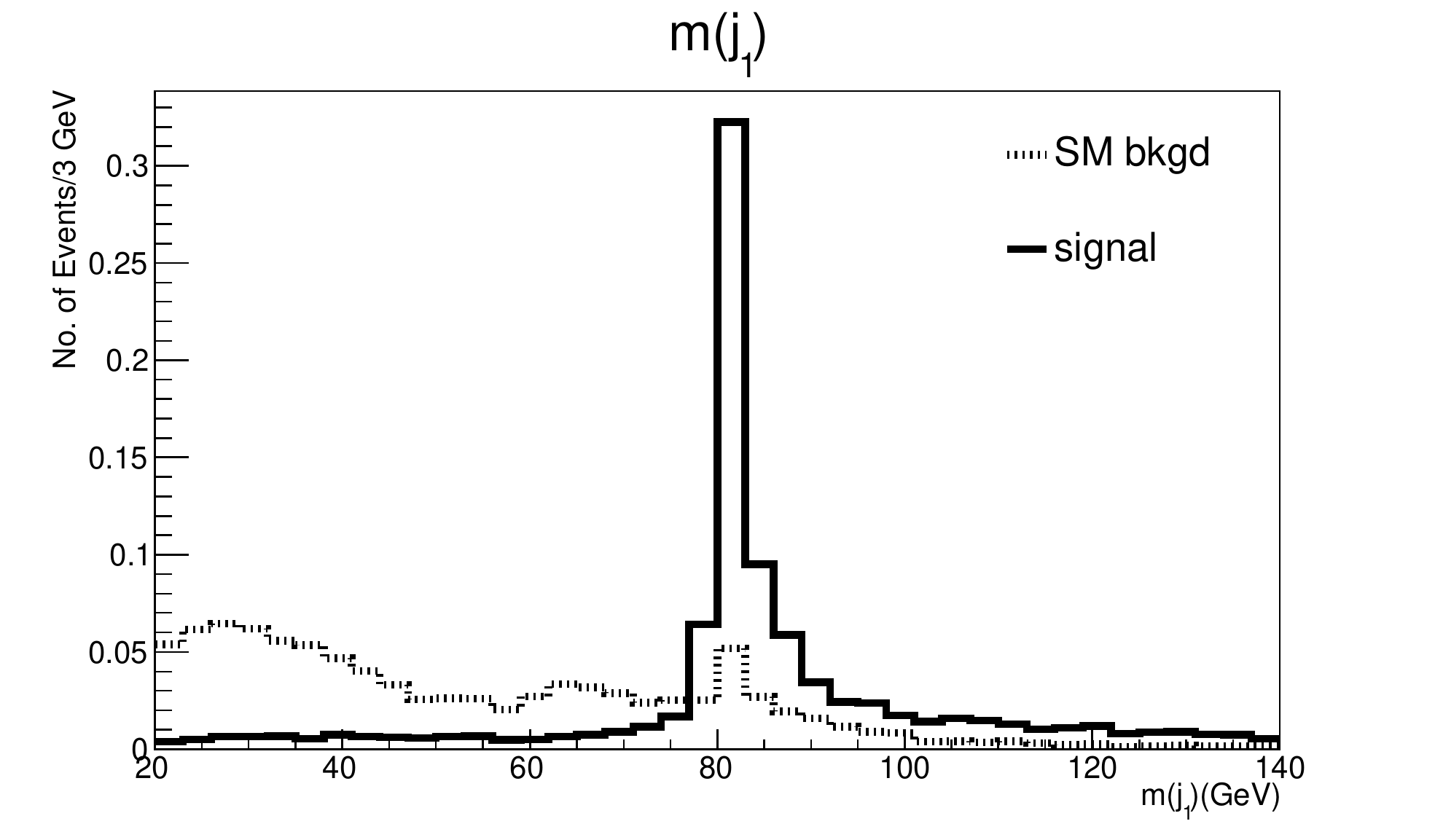} \includegraphics[width=0.45 \columnwidth]{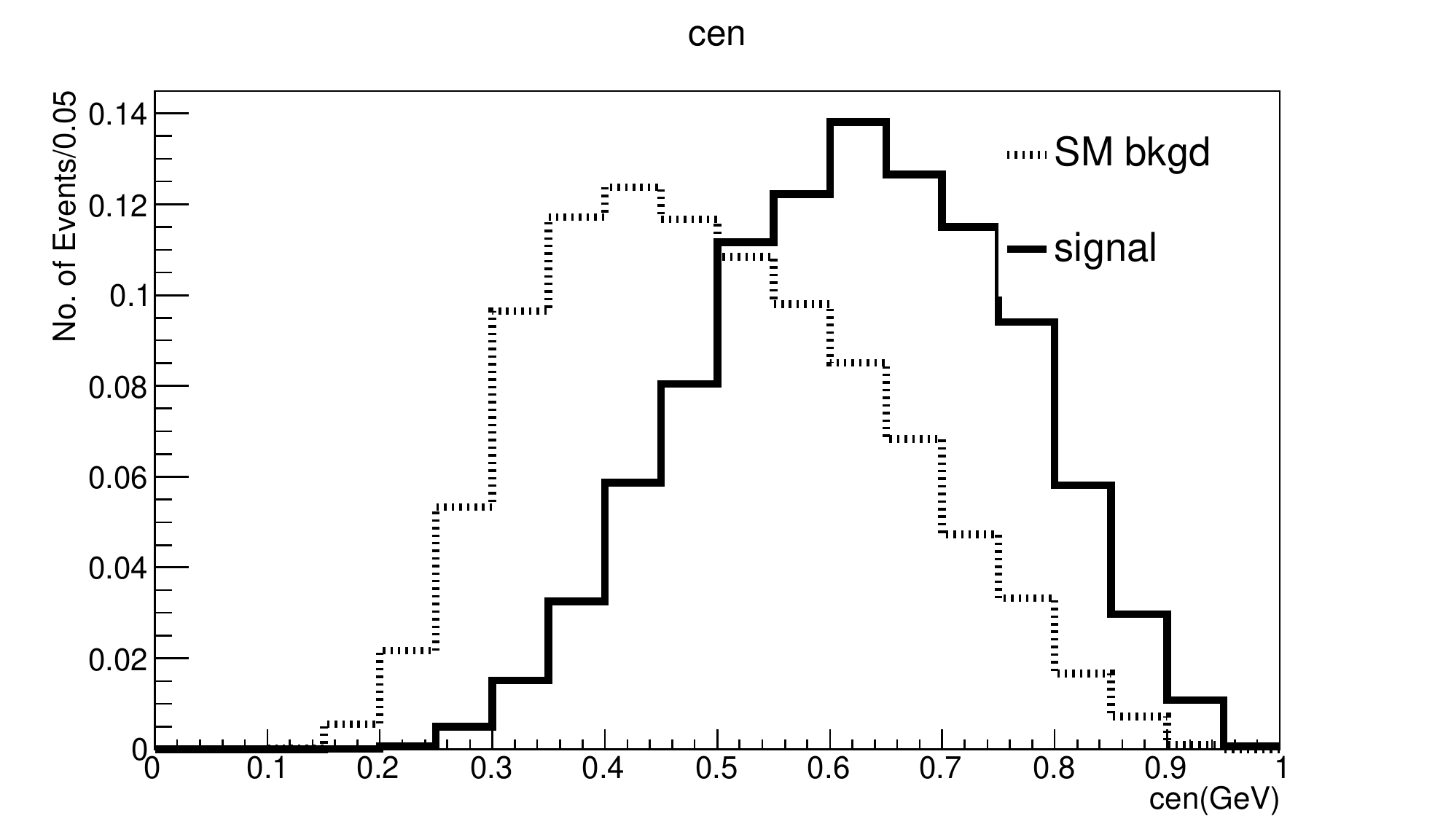}
\caption{The observables showing the event shape and small jets for signal and background discrimination analysis are displayed, which can be obtained without using reconstruction procedure.
\label{obs-no-rec}}
\end{center}
\end{figure}

Below we roughly describe the crucial observables which can be used to distinguish signal and background. We can define the kinematic observables into two categories: observables without reconstruction and observables with reconstruction. The observables without reconstruction include observables which can be directly extracted from jets in the final state. For example, the transverse momentum and invariant mass of leading two jets can be obtained once we specify the jet clustering algorithms. Event shape observables, like the $H_t$, $\hat s$, centrality, and sphericity, can also be computed. As demonstrated in Fig. \ref{obs-no-rec}, where the transverse momentum and invariant mass of leading two jets, the event shape variables, $H_t$, ${\hat s}$, and centrality are shown.

\begin{figure}[!htb]
\begin{center}
\includegraphics[width=0.45 \columnwidth]{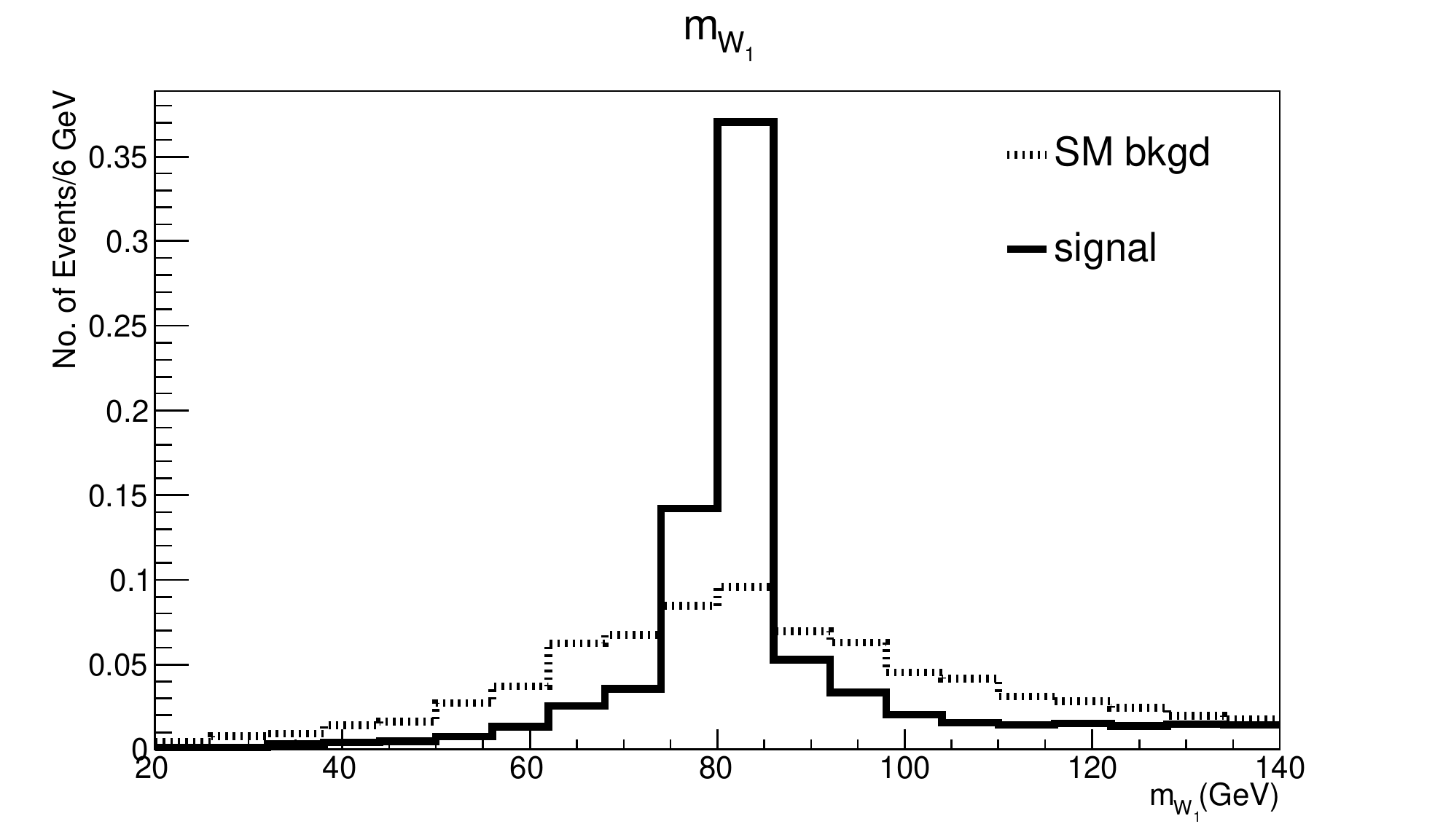} \includegraphics[width=0.45 \columnwidth]{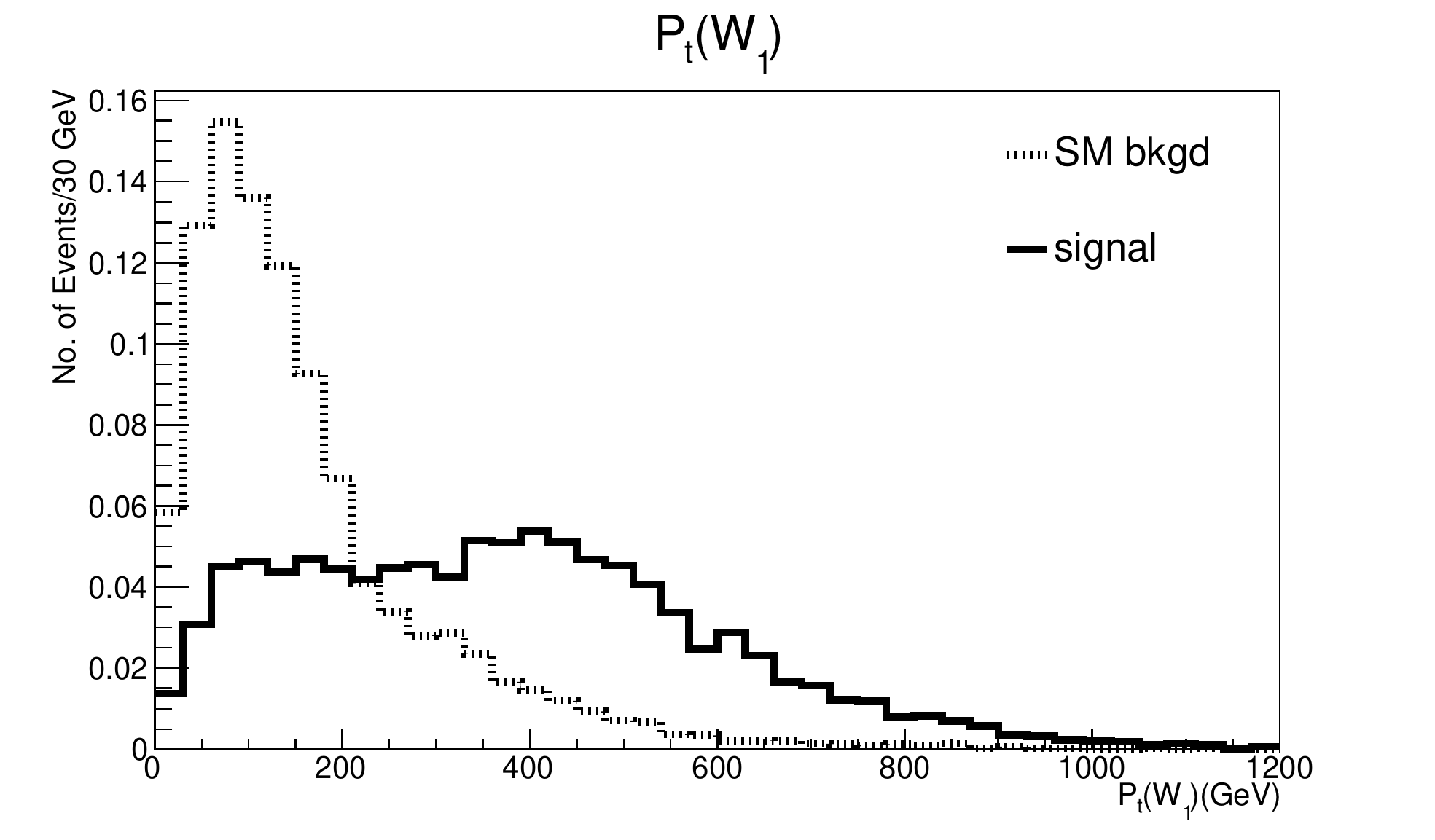} \\
\includegraphics[width=0.45 \columnwidth]{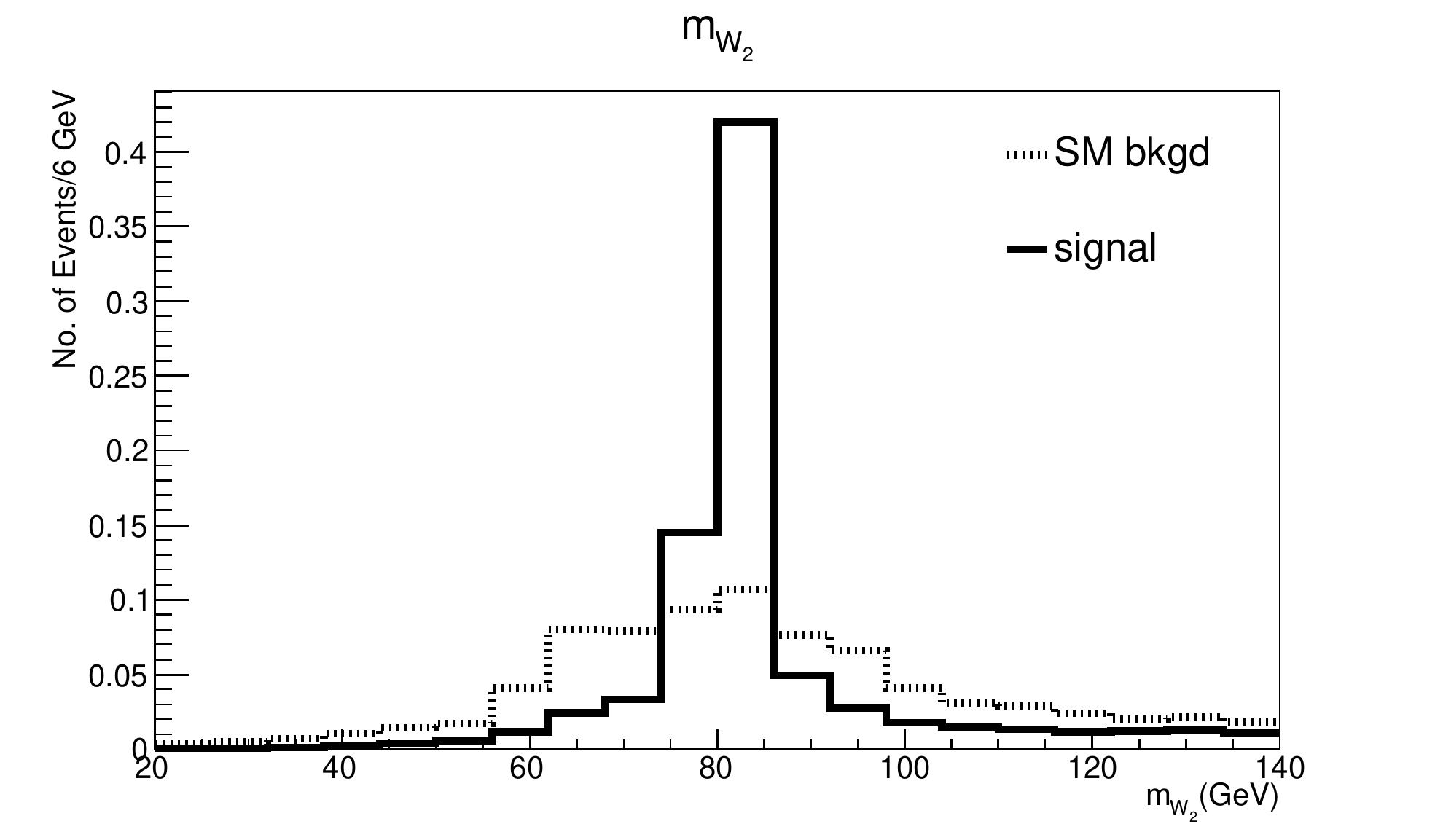} \includegraphics[width=0.45 \columnwidth]{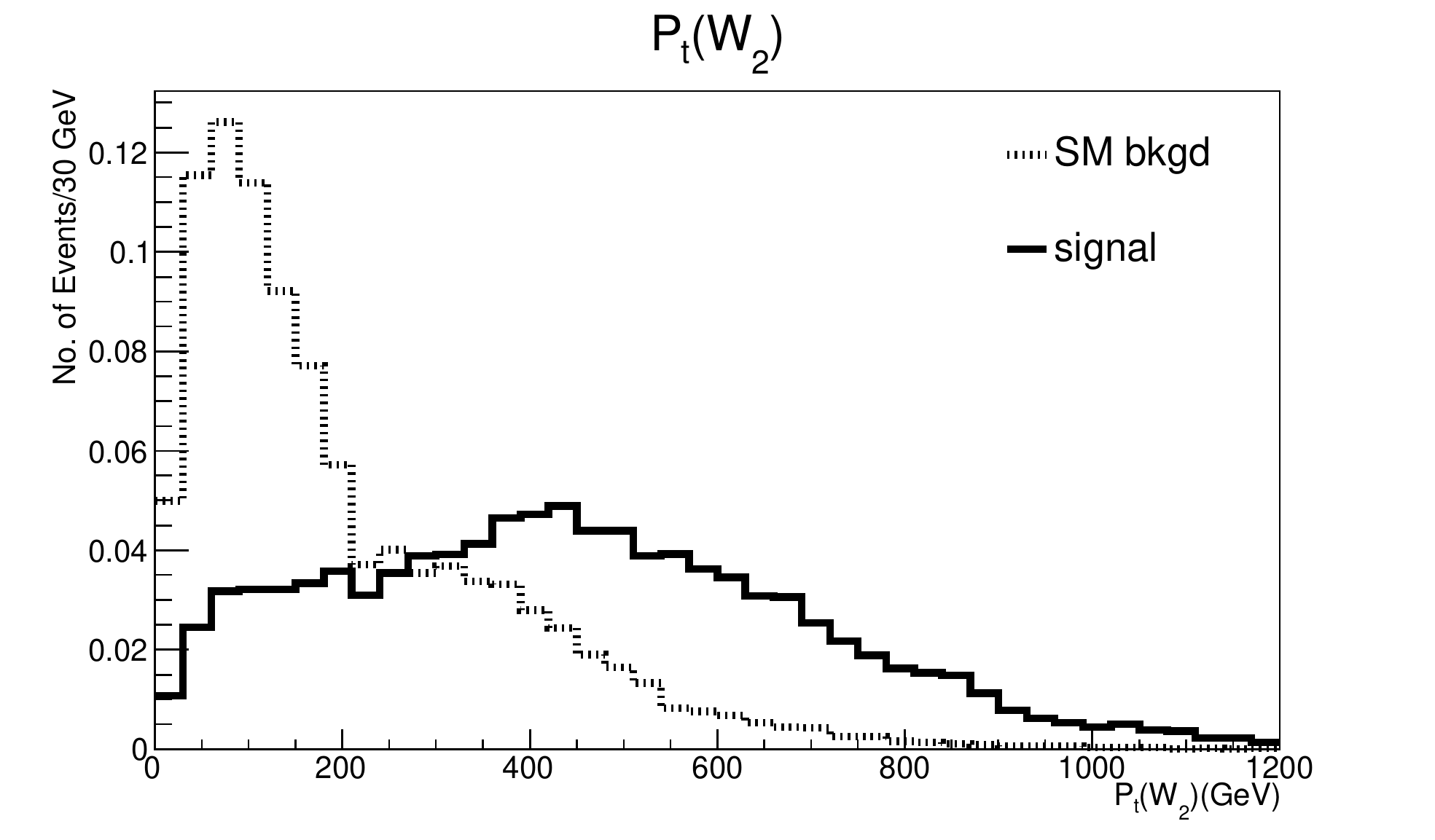}\\
\includegraphics[width=0.45\columnwidth]{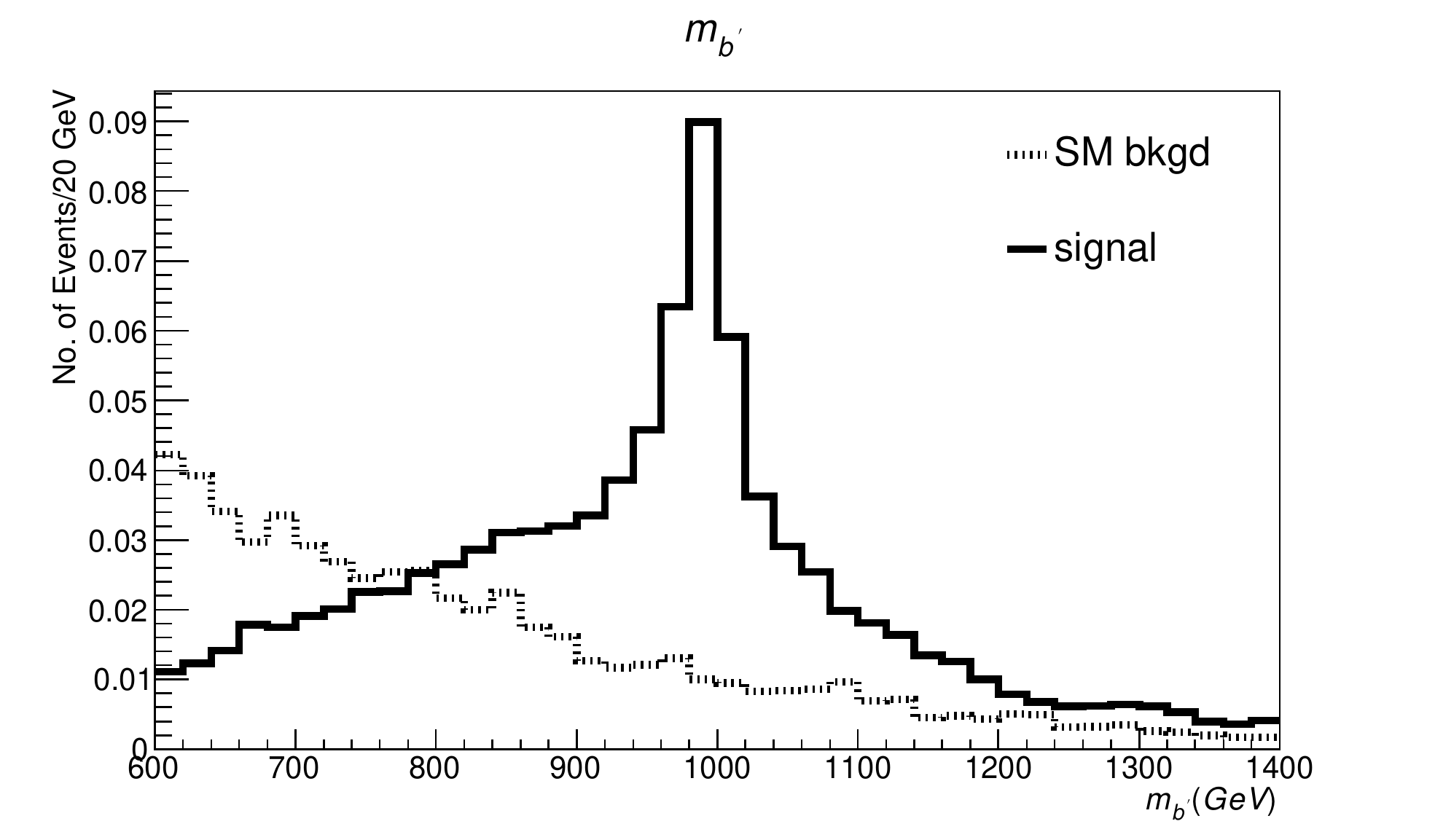}  \includegraphics[width=0.45\columnwidth]{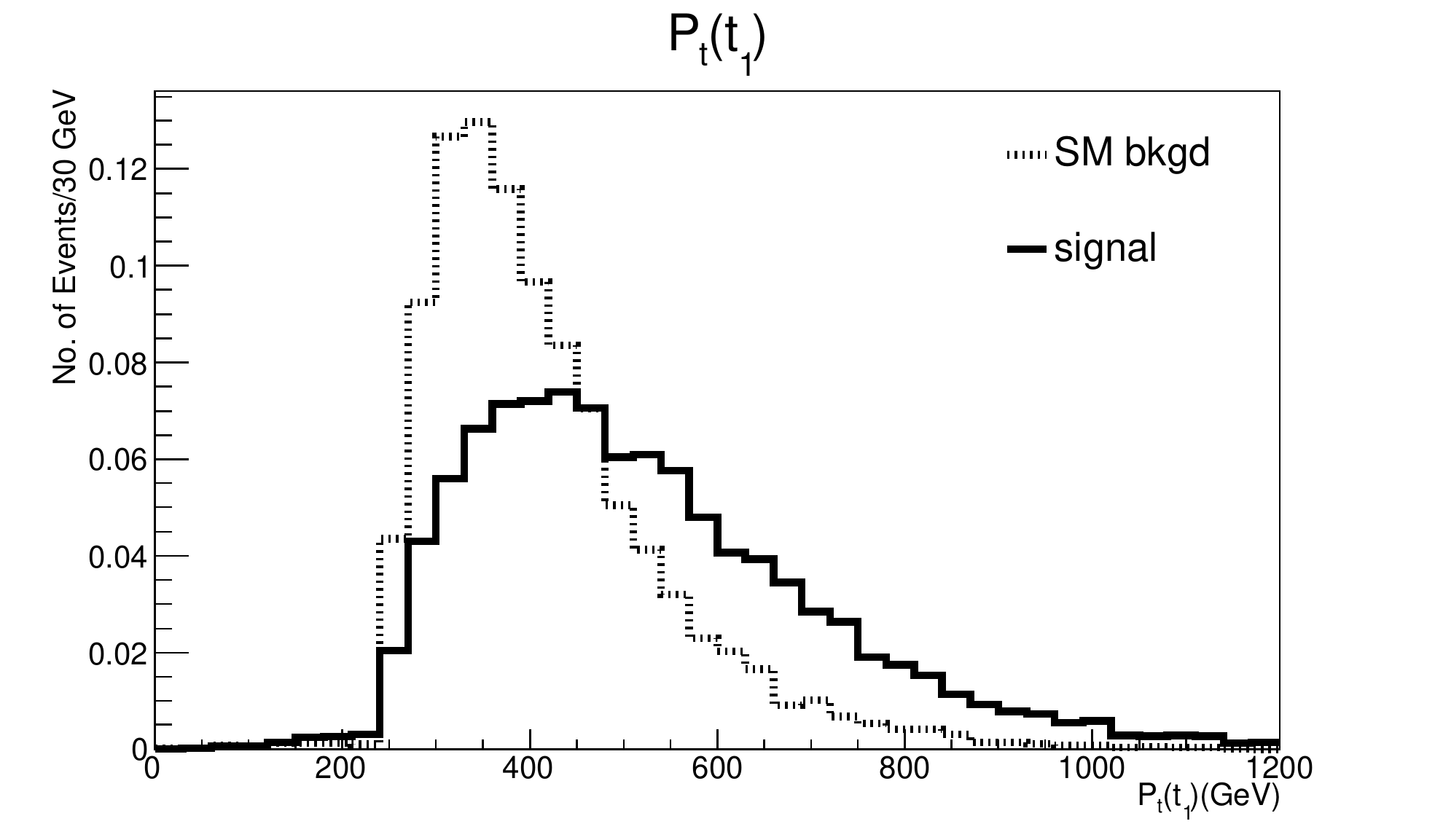}
\caption{The observables for intrinsic theoretical parameters and phase space of signals are displayed, which can be obtained by using our proposed reconstruction procedure.
\label{obs-have-rec}}
\end{center}
\end{figure}

The second type of observables are those which can only be obtained after reconstruction. For example, the reconstructed masses, transverse momenta, and $\eta$s of W bosons, top quarks and $b^{\prime}$ can only be obtained after we can identify all physics objects in term of our reconstruction procedure. In Fig. \ref{obs-have-rec}, we show the most useful and important observables which can help to discriminate signals and backgrounds, like the transverse momenta of reconstructed W bosons and top quarks, the transverse momentum of reconstructed $b^{\prime}$.

\begin{figure}[htbp]
 \hspace{0.0cm}
\includegraphics[width=6.5in]{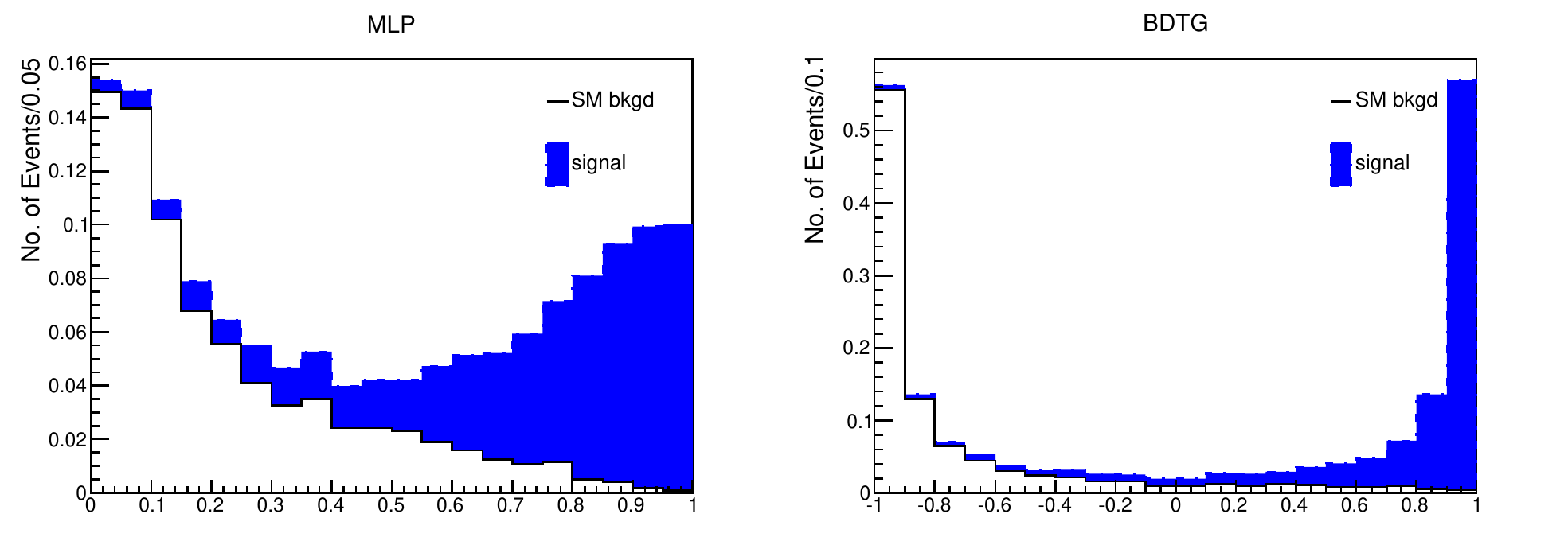}
\caption{The signal and background discriminations in the MLP NN method and BDT method are demonstrated.
}
\label{figmva}
\end{figure}

For signal events, all these two types of observables are intrinsically correlated to the mass parameter of $b^\prime$. In contrast, for the background events there is no such a correlation.  By utlizing these observables and this correlation, we use the package TMVA to perform the training process and then apply the determined weights for each events which have not seen by the training process. We have applied both MLP neural network method and the boosted decision tree method \cite{Roe:2004na,Yang:2005nz,Yang:2007pb}. The discriminant distributions for these two MVA methods are provided in Fig. \ref{figmva}, which clearly demonstrate the discriminant analysis indeed works. We have also used the cut based method and observed that the MVA methods can optimize the signal and background discrimination better and improve the significance by a factor $100\%$ or so, similar to the observation in our previous work \cite{Yang:2011jk} where the heavy charged Higgs boson search was studied. After using the MVA cuts, as demonstrated in Fig. \ref{figmass}, we can see the mass bump of reconstructed $b^{\prime}$ clearly standing out from the SM background.

\subsection{Sensitivity of the LHC to $b^\prime$}
We use the analysis presented above to other values of $m_{b^\prime}$ from 800 GeV to 1500 GeV, where top quarks from $b^\prime$ decay can be either intermediately or highly boosted, and we arrive at the sensitivity given in Table \ref{tagmodes-bkgd}. We observe that when the $b^\prime$ is heavier its production rate becomes smaller, which leads to a smaller significance.
\begin{table}[th]
\begin{center}%
\begin{tabular}
[c]{|c|c|c|c|c|c|c|c|c|}\hline
$m_b^\prime$ & 0.8   &  0.9  & 1.0 & 1.1   & 1.2   & 1.3 & 1.4   & 1.5  \\ \hline
Signal & 1316 & 708 & 356 & 209 & 108 & 58 & 38 &  14 \\ \hline
Background & 2560 & 1797 & 992 & 866 & 346 & 291 & 184 & 87  \\ \hline
$\frac{S}{\sqrt{B}}$ & 26.0 & 16.7 & 11.3 & 7.1 & 5.8 & 3.4 & 2.8 & 1.5 \\ \hline
\end{tabular}
\end{center}
\caption{The significance achieved in our analysis is tabulated by varying the $b^\prime$ mass from 0.8 TeV to 1.5 TeV. The optimized cone parameters for both W boson and top quark jets are provided in the first row. The number of signal and background are normalized to be 200 fb$^{-1}$ with the collision energy 14 TeV. }%
\label{tagmodes-bkgd}%
\end{table}

In Table (\ref{table:scanning}), we examine the upper bound on the cross section of $pp \to b^\prime {\bar b^\prime}$, where the significance $\sigma=2.5$ is used to compute the exclusion bound.
\begin{table}[th]
\begin{center}%
\begin{tabular}
[c]{|c|c|c|c|c|c|c|c|c|}\hline
$m_{b^\prime}$ (TeV) & $0.8$ &  $0.9$  & $1$ & $1.1$ & $1.2$ &$1.3$ &$1.4$ &$1.5$ \\ \hline
$\sigma$ (fb) & $288.2$ & $137.2$ & $69.0$ & $36.2$ & $19.7$  & $11.0$ & $6.3$ & $3.7$  \\ \hline \hline
$\frac{S}{\sqrt{B}}$ (with two $b$ taggings and TMVA) & $26.0$ & $16.7$ & $11.3$ & $7.1$ & $5.8$ & $3.4$ & $2.8$ & $1.5$ \\ \hline
lower bound on $\sigma(fb)$ & $22.2$ & $16.4$ & $12.3$ & $10.2$  & $6.8$  & $6.5$  & $4.5$ &  $4.9$ \\ \hline \hline
\end{tabular}
\end{center}
\caption{The significance and sensitivity of LHC for $b^\prime$ production are shown, where we assume the total integrated luminosity as 200 fb$^{-1}$. We use $\frac{S}{\sqrt{B}}=2.5$ to define the exclusion upper bound on the cross section.}%
\label{table:scanning}%
\end{table}

\begin{figure}[htbp]
 \hspace{0.0cm}
\includegraphics[width=6.5in]{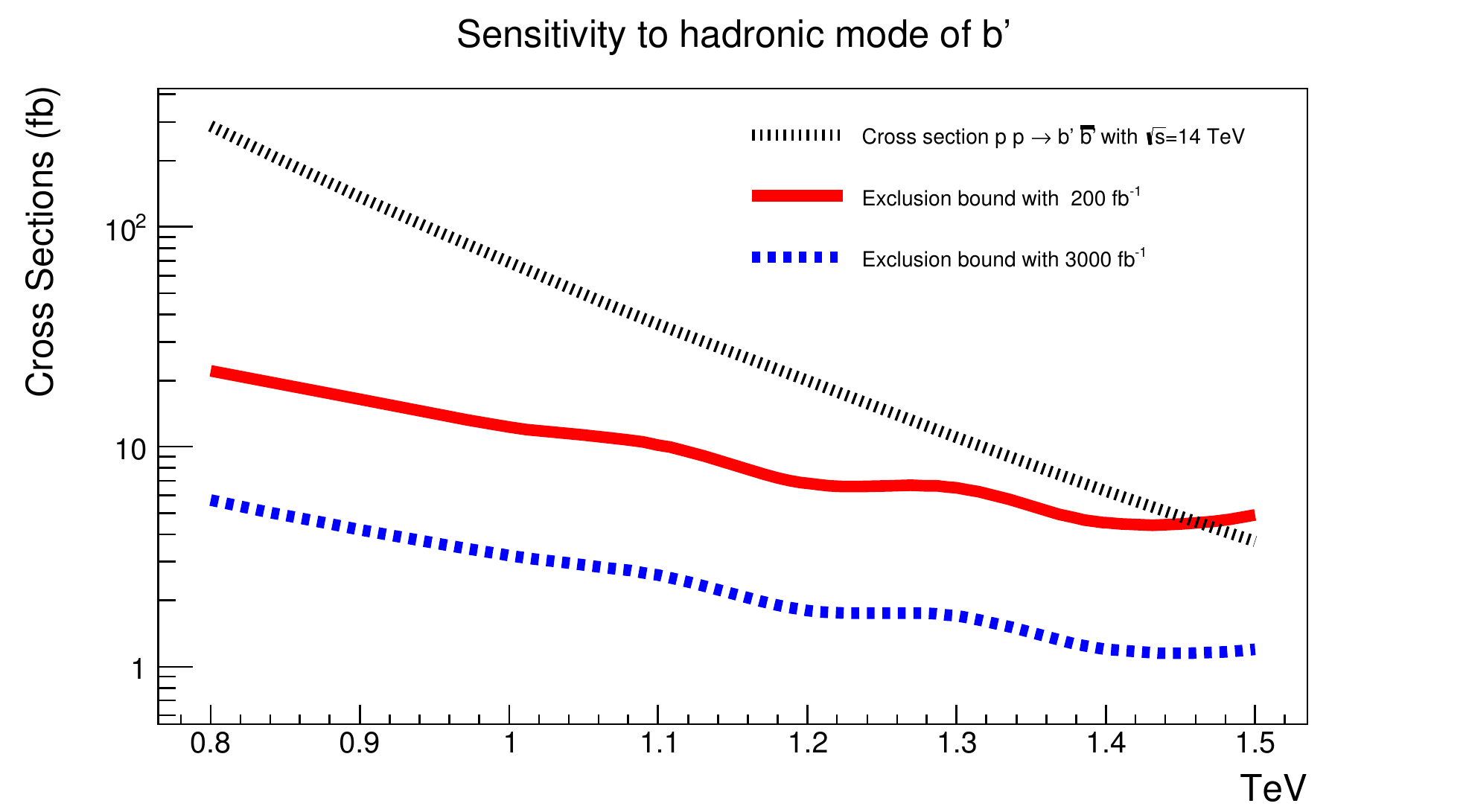}
\caption{The sensitivity to the signal of hadronic $b^{\prime}$ mode at LHC 14 TeV is shown. The results for 3000 fb$^{-1}$ dataset is obtained by directly scaling up the results of 200 fb$^{-1}$.
}
\label{figsens}
\end{figure}

In Fig. \ref{figsens}, we present the sensitivity of the LHC to the hadronic $b^\prime$ mode with integrated luminosity 200 $fb^{-1}$ and 3000 fb$^{-1}$, respectively. From this plot, we can estimate that with 3000 fb$^{-1}$ dataset just using the hadronic $b^\prime$ mode, we can either find or rule out a $b^\prime$ up to 1.8-2.0 TeV or so. Compared with the semileptonic modes analyzed in \cite{Skiba}, it is expected that the hadronic mode demands more luminosity due to the large SM background events, similar to the case of the hadronic mode of $t{\bar t}$.

\section{Discussions and Conclusions}

In this paper, we have studied the full hadronic mode for the process $pp \rightarrow b'b' \rightarrow tW^-\bar{t}W^+$ at the LHC 14 TeV collision.
The main task is to reconstruct the $b^{\prime}$ from the large combinatorics and to suppress the huge QCD and $t{\bar t} + $ jets background events, where
we have found that $b$ taggings are essential to suppress background events with large jet multiplicity from QCD. By using the top-tagger and W-tagger and some comprehensible cuts, we propose a full reconstruction procedure and demonstrate that if we could reconstruct the most important parameters of the signal, like the mass of $b^\prime$ and transverse momentum of W bosons, etc., the hadronic mode of $b^\prime$ could be feasible for the future LHC runs.

The current work can be directly extended for the higher energy collisions (say 100 TeV collisions). Obviously, for the signal with a fixed mass (say $m_b^\prime=10$ TeV),  the signal will be enhanced due to the increase of collision energy and the corresponding large enhancement in the gluon fluxes. Nonetheless, more background processes may become dominant, for example, the multiple W boson final states, $t{\bar t} + W$, $t{\bar t} + WW$, $t{\bar t} t {\bar t}$, and $t W$. Some of these backgrounds could have significantly large enhancement in cross sections, say $t{\bar t} t {\bar t}$ and $t{\bar t} h$. Moreover, the collimated W bosons from EW showers \cite{Christiansen:2014kba,Krauss:2014yaa} in energetic $b$ jets might fake the top quark taggers to some degree, which might be crucial for high energy collisions. Another challenging issue for 100 TeV collisions study is to model the multijet background events from the SM. We leave a detailed analysis for high energy collisions in our future works.

To project the sensitivity for $b^\prime$ at higher energy collisions, it might be useful to take into account leptonic modes. If so, a top tagger with leptonic modes should be useful \cite{Stoplep}. When combining both leptonic and hadronic modes, we expect that higher exclusion bounds for $b^\prime$ can be achieved or a better significance can be obtained if a $b^\prime$ is there.

We have not included the pileup effects to the top and W taggers here, which is necessary for a more realistic analysis at either LHC future runs or future high energy collisions. As demonstrated in the reference \cite{Anderson:2013kxz}, the pileup effects might decrease the significance to a certain degree. Another interesting thing for high energy collisions is that when a top quark is very highly boosted (say $P_t > 5$ TeV), the current top tagger should be improved as demonstrated in \cite{Schaetzel:2013vka} by taking into account additional information from tracker system, or it might be improved from hardwares, say by increasing the granularity of detectors from $0.1\times 0.1$ to $0.01 \times 0.01$. Obviously, studying boosted physics objects and their taggings at high energy can help us in the detector designs for future high energy collisions.

\acknowledgments

Q.S Yan would like to thank Qing-Hong Cao, Michelangelo L. Mangano, Li-Lin Yang for useful discussions and comments. This work was supported in part by the National Science Foundation of China under Grant Nos. 11175251, 11205023, 11375248 and by the Natural Science Foundation of Dalian under No. 2013J21DW001.

\end{document}